\documentclass[hidelinks,journal]{IEEEtran}
\usepackage{latexsym}
\usepackage{graphicx}
\usepackage{amsfonts,amssymb,amsmath}

\usepackage{hyperref}

\usepackage{subcaption}
\usepackage{listings}
\usepackage{xcolor}

\lstdefinelanguage{PowerShell}
{
    morekeywords={Get-ChildItem, Where-Object, GetValue, eq},
    morestring=[b]',
    morestring=[b]"
}

\lstdefinestyle{powershellstyle}{
    language=PowerShell,
    basicstyle=\ttfamily,
    keywordstyle=\color{blue},
    commentstyle=\color{green},
    stringstyle=\color{red},
    numbers=left,
    numberstyle=\tiny,
    numbersep=5pt,
    breaklines=true,
    breakatwhitespace=true,
    showstringspaces=false,
    frame=single,
    rulecolor=\color{black},
    tabsize=4,
    backgroundcolor=\color{gray!10},
}

\def\BibTeX{{\rm B\kern-.05em{\sc i\kern-.025em b}\kern-.08em T\kern-.1667em\lower.7ex\hbox{E}\kern-.125emX}}
\begin{document}

\title{A Bird-Eye view on DNA Storage Simulators}

\author{Sanket Doshi, Mihir Gohel and Manish K. Gupta\thanks{Authors are with the Dhirubhai Ambani Institute of Information Communication Technology, Gandhinagar, Gujarat 382007, India. \\Email: sanbdoshi@gmail.com, mh.dgohel@gmail.com, \\mankg@guptalab.org}}



\maketitle

\begin{abstract}
In the current world due to the huge demand for storage, DNA-based storage solution sounds quite promising because of their longevity, low power consumption, and high capacity. However in real life storing data in the form of DNA is quite expensive, and challenging. Therefore researchers and developers develop such kind of software that helps simulate real-life DNA storage without worrying about the cost. This paper aims to review some of the software that performs DNA storage simulations in different domains. The paper also explains the core concepts such as synthesis, sequencing, clustering, reconstruction, GC window, K-mer window, etc and some overview on existing algorithms. Further, we present  3 different softwares on the basis of domain, implementation techniques, and customer/commercial usability.
\end{abstract}



\section{INTRODUCTION}\label{sec:1}
Data storage is a big problem in today's world. More and more, data is generated on a daily basis, approximately 27 zeta-bytes of increment in 2024 \cite{yearwiseincrementofdatastorage}. Lots of data centers/storage capacity is needed to store this much amount of data. Due to which an alternative should be invented in order to store this big data. And also the data that is stored on our current storage hardware are prone to decay \cite{tabatabaei2015rewritable}. So, the alternative is to store this big data into DNA \cite{church2012next}. DNA gets stored in a cool, dark environment without the need for electricity, unlike conventional storage devices. \cite{shah2013dnacloud}. Theoretically, in one gram of synthetic DNA molecule, 455 exabytes of data can be stored and retrieved \cite{church2012next}. DNA storage ensures longevity, low power, and capacity \cite{hoshika2019hachimoji}. In today's world it costs around \$3500 for storing 1MB into DNA \cite{costDNAStorage}. So in-spite of being low-power it has a huge cost and it is due to synthesis and sequencing processes.

Synthesis and sequencing processes result in a huge amount of cost in the entire DNA storage process because special reagents and instruments are used. Also, the machines that are involved in converting digital data into DNA and the DNA back to digital data are very expensive. Hence instead of ensuring low power, the DNA storage process is expensive. Due to this high amount of cost, it is not possible to test out new algorithms for DNA storage but we can somehow mimic the DNA storage process into some sort of software called DNA storage simulator to test the newly developed algorithms. DNA storage simulators are build around different domains, for example some of them are sequencing specific whereas some other are storage specific (see Section \ref{sec:3} for more details). 

This review will discuss the whole process involved into DNA storage, such as encoding, synthesis, sequencing, clustering, reconstruction. All of which can be seen in Section \ref{sec:2}. Further on, in Section \ref{sec:3} discussion will be carried on to the core part, that is DNA storage simulators, where three different domain specific DNA storage simulators are chosen. All different algorithms associated with sequencing, synthesis, clustering, reconstruction, PCR process, etc, are mentioned for each of these simulators. Simulations for all these simulators, along with how to perform these simulations are also mentioned. And lastly probable errors and improvements for each simulators has been discussed. Section \ref{sec:4} provides a crisp comparison between these simulators. And at the very end in Section \ref{sec:5}, \ref{sec:6}, and \ref{sec:7} discusses about technology limitations, future scope, and conclusion respectively.


\section{DNA storage process}\label{sec:2}
\begin{figure*}[h]
	\centering
	\includegraphics[scale=0.65]{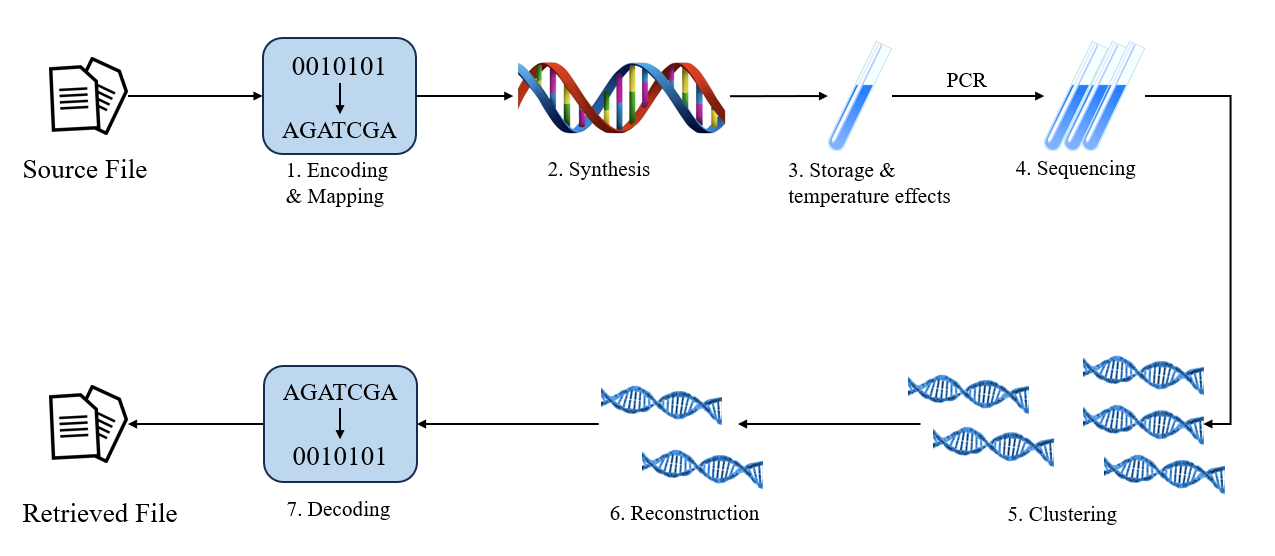}
\caption{Workflow of the entire DNA storage process. First a source string (in this case binary string) is broken down into chunks of same length. Then, two stages of encoding process happens, mapping of binary to quaternary (i.e., ACGT) format followed by adding extra bits of information for error correction purpose, by using LDPC codes, Reed Solomon codes, etc. After that these DNA chunks (oligos) are stored into synthetic DNA molecules using DNA synthesizers and those synthetic DNA molecules are stored under certain medium. After that, at the time of reading process, multiple copies of each DNA strands are generated using PCR amplification to easily retrieve necessary information. Then, clustering algorithm is applied to group identical oligos and at the last by using reconstruction followed by decoding
process (which is reverse of encoding), original binary file is produced.}
\label{DNAstorageprocess}
\end{figure*}

The entire DNA storage process can be primarily divided into 7 steps: encoding, synthesis, storage host, sequencing, clustering, reconstruction, and decoding (see Fig.~({\ref{DNAstorageprocess}})). First a source string is broken down into chunks of same length. Then, extra bits of information is added for error correction, followed by mapping those final chunks to quaternary format. After that oligos are stored under synthetic DNA strands by synthesis process and those DNA strands are stored under cold environment. At the time of information retrieval (reading process), multiple copies of each DNA strands are generated using PCR amplification. Then, clustering is used to group similar oligos and at last by using reconstruction followed by decoding process (which is reverse of encoding), output file is produced. Complete analysis of each steps is provided in this section only. For further details, interested readers may refer to \cite{10yearsofnaturaldatastorage}.

\subsection{Encoding}
Encoding \cite{meiser2020reading} is used to convert the provided data (in any format) to a format that is suitable for DNA storage (i.e., ACGT format, Quaternary format). Encoding can be broken down into two parts. First, extra bits of information are added in order to recover the original information with more accuracy at the time of decoding. Various algorithms such as LDPC codes, Reed Solomon codes,... are used for this purpose. Those algorithms are known as error-correcting codes. Secondly, using hash tables binary to ACGT mapping is performed. Mainly non-linear ternary codes \cite{limbachiya2015optimal} are used here.

\subsection{Synthesis}
Strings of ACGT format are stored in the actual DNA strands by using a synthesis process \cite{meiser2020reading}. This can be done with the help of machines called DNA array synthesizers. It is more beneficial to synthetically construct DNA rather than changing the natural DNA sequences using chemical enzymes. This method works on Oligo-Pools (they are a mixture of more than thousand individually designed Poly Nucleotides). This Oligo-pools based synthesis method is very effective as it provides very low error rates at a very reasonable cost. Traditionally, array-based synthesis technologies were used which were built on photolithographic masks. This can be done by protecting and de-protecting the nucleotides (consisting of a sugar molecule, a nitrogen-containing base, and a phosphate group) of a DNA with the help of directing the UV light towards it. In the present day, instead of UV lights, thousands of microelectrodes are used in electrochemical reactions for the protection/de-protection process. However, the synthesis process has one problem. When the given string is stored in DNA strands, the information does not remain present altogether as the DNA strands are randomly arranged. Therefore, it can not be determined which DNA strand belongs to which file. This problem can be solved by encapsulating the DNA file in a silica molecule.

\subsection{Sequencing}
Sequencing \cite{meiser2020reading} is the process in which the actual DNA is converted back to digital information in computer-readable form. Polymerase Chain Reaction (PCR) is the method used in laboratories to amplify the actual DNA strands. After the amplification process, it is quite easy to retrieve the required information from the DNA strands with minimal loss. However, accessing a particular data file from the DNA strands can be quite difficult as the DNA strands can be randomly arranged in sequences without keeping track of the data file they belong to. Suppose DNA strands contain two different data files, file 1 and file 2. The required information file is file 1. During the amplification step, it might be possible that the data of file 2 has been partially amplified. Hence the data of file 2 might be lost during the PCR process. This problem can be solved by encapsulating the DNA file in a silica molecule and by using PCR primer\footnote{A PCR Primer can be understood as a comb-like structure i.e. a single-stranded DNA sequence. This primer is responsible for amplifying the DNA sequence of the data file.}.
Here, each data file is associated with a PCR primer, which is labeled as fluorescent or magnetic particles. With each PCR primer, a bar code is associated. Hence the required file can easily be retrieved without losing any other files\footnote{https://news.mit.edu/2021/dna-data-storage-0610}. Currently, the sequencing process is performed by a more commonly used platform called “Illumina”. There are multiple models that are available as per the demands of the user. Some of these models are NextSeq Series, iSeq 100 series, etc. Currently, Illumina dominates the majority of the work platform that is present for the sequencing process.

\subsection{Clustering}
Clustering uses the concept of K-means clustering used in ML problems. Here number of clusters is assumed beforehand and then the algorithm is executed. Nowadays clustering is done by K-means clustering along with dynamically updated hash index (DUHI) \cite{wang2023duhi}
method. Which involves indexing along with ML concepts to handle the problem better. This solution helps in reducing the redundancy within the clusters and increases reconstruction rates massively. For more advanced and latest clustering algorithms interested readers may refer to \cite{clusteringbillionsofreads} \cite{clover}.

\subsection{Reconstruction}
Let us have a n-length input string x and pass it over to a channel of deletion-substitution-insertion t-times. Corresponding to this it’ll generate t-traces $y_{1},y_{2},…,y_{t}$. Now this t-traces will be mapped to a n-length string $\hat{x}$. It means taking $y_{1},y_{2},…,y_{t}$ as model parameters and producing an output $\hat{x}$. Ideally, this $\hat{x}$ should be equal to x. The main goal for this algorithm is to minimize the edit distance between x and $\hat{x}$. So, this turns out to be an ML problem \cite{sabary2023reconstruction}.

\subsubsection{Issue}
The above reconstruction algorithm assumes that every strand in a cluster is a noisy copy that originates from the same reference and hence it believes that they contribute equally to the reconstruction process. However, this assumption is not always true. Due to DNA rearrangement and fragmentation that occurs during the DNA storage process, the cluster of noisy copies of DNA strands also contain outlier sequences. These outlier sequences are called contaminated clusters. Therefore, the simple reconstruction process can not be applied here due to contaminated clusters. Because each strand within a cluster now makes a different contribution to the reconstruction of the baseline strand \cite{sabary2023reconstruction}.

\subsubsection{DNN Based Reconstruction Method}
A deep neural network based reconstruction method \cite{qin2022robust} is robust to the contaminated clusters which contain outlier sequences and noisy reads with insertion/deletion/substitution errors. This is the first multi read reconstruction method in which all the strands from a cluster contribute unequally to the reconstruction of the baseline strand. This neural network structure has approximately 2.5 M parameters. If N is cluster size, in which er is the number of erroneous reads originating from the same reference and cs is number of contaminated sequences then, this algorithm assumes $N = er + cs$. And based on this assumption this algorithm takes n sequences as input and produces as output $\hat{x}$ (which ideally should be equal to the reference string).

\subsection{Decoding}
The received 'ACGT' format from the reconstruction stage should be converted to digital data. This conversion process is known as decoding \cite{meiser2020reading}. This process is a reverse of the encoding process. The decoder accepts a text file or a .fasta file, which contains the long string in ACGT format. Then reverse process of conversion from ACGT to binary happens using the hash tables, that were used in the encoding. And then finally reed solomon codes/LDPC codes is applied to remove redundant bits.


\section{DNA Storage Simulators}\label{sec:3}
\begin{figure*}[h]
	\centering
	\includegraphics[scale=0.55]{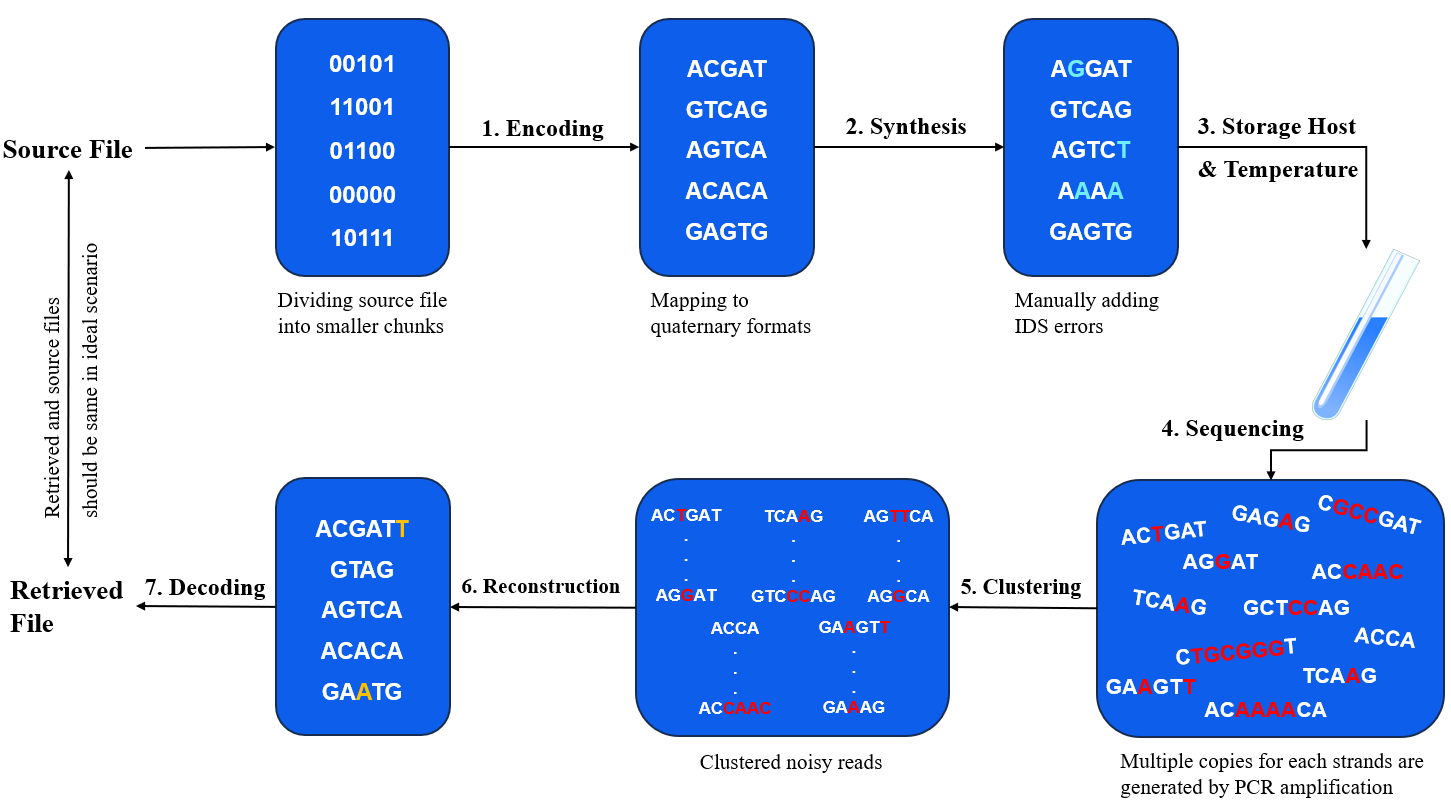}
\caption{This figure shows workflow of an ideal DNA storage simulator. It mainly considers 7 steps: first, it takes a source file as an input, breaks it into same sized chunks, maps it to quaternary code words (i.e., ACGT) and add some extra redundancy for error correction. Then, simulator adds some IDS errors which occur during actual synthesis process. Storage and temperature effects also needs to be thoroughly studied. After that, it mimics the PCR process in order to read necessary information files. At last clustering, reconstruction and decoding steps are used to produce the desired output file.}
\label{Idealsimulatorworkflow}
\end{figure*}

\begin{figure}[h]
	\centering
	\includegraphics[scale=0.60]{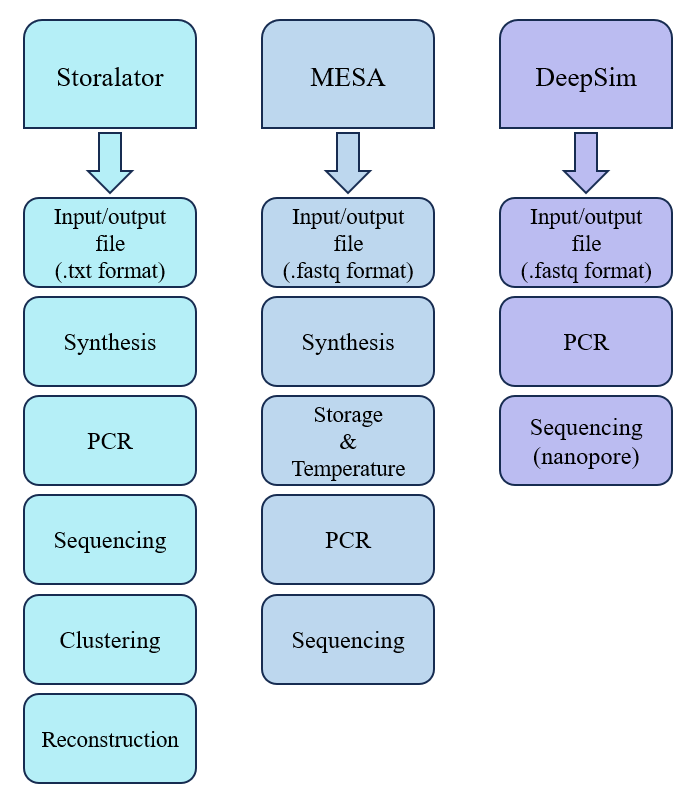}
\caption{Block diagrams of basic DNA storage processes that are linked with Storalator, MESA, and DeepSimulator softwares respectively.}
\label{Simulatorsblockdiagram}
\end{figure}

As discussed in Section \ref{sec:1}, purpose of DNA storage simulators is to test different algorithms, which can then be used in the actual DNA storage process in order to achieve more efficiency and cost reduction. The simulator helps in figuring out the impact of encoding/decoding algorithms via the statistical behavior of the noise signal and can also simulate the PCR process without worrying about the costs involved. There are some existing simulators in the market such as DNA storage error simulator \cite{dnastorageerrorsimulator}, PBSIM2 \cite{pbsim2}, PBSIM3 \cite{pbsim3}, NanoSim \cite{nanosim}, Badread \cite{badread}, Storalator \cite{storalatorpaper}, MESA \cite{schwarz2020mesa}, and DeepSimulator \cite{li2018deepsimulator}. We have analyzed these three simulators (their software links are provided in references): Storalator \cite{storalatormainsoftware}, Mesa \cite{mesamainsoftware}, and DeepSimulator \cite{deepsimulatormainsoftware} due to their diversity in performing DNA storage simulations in different domains and also they include different steps associated with DNA storage process. Based on the facts and by keeping DNA storage process as reference, we have prepared workflow of an ideal simulator (see Fig.~({\ref{Idealsimulatorworkflow}})). All three simulators do not have any features for encoding and decoding. Apart from that, storalator has all the parameters except storage and temperature effects, mesa does not contain clustering and reconstruction algorithms, and deepsimulator is only based on nanopore sequencing technology (Fig.~({\ref{Simulatorsblockdiagram}}) represents the high level block diagram of above simulators). In the upcoming subsections detailed analysis of each simulators is provided.

\subsection{Storalator}
\begin{figure*}[h]
	\centering
	\includegraphics[scale=0.7]{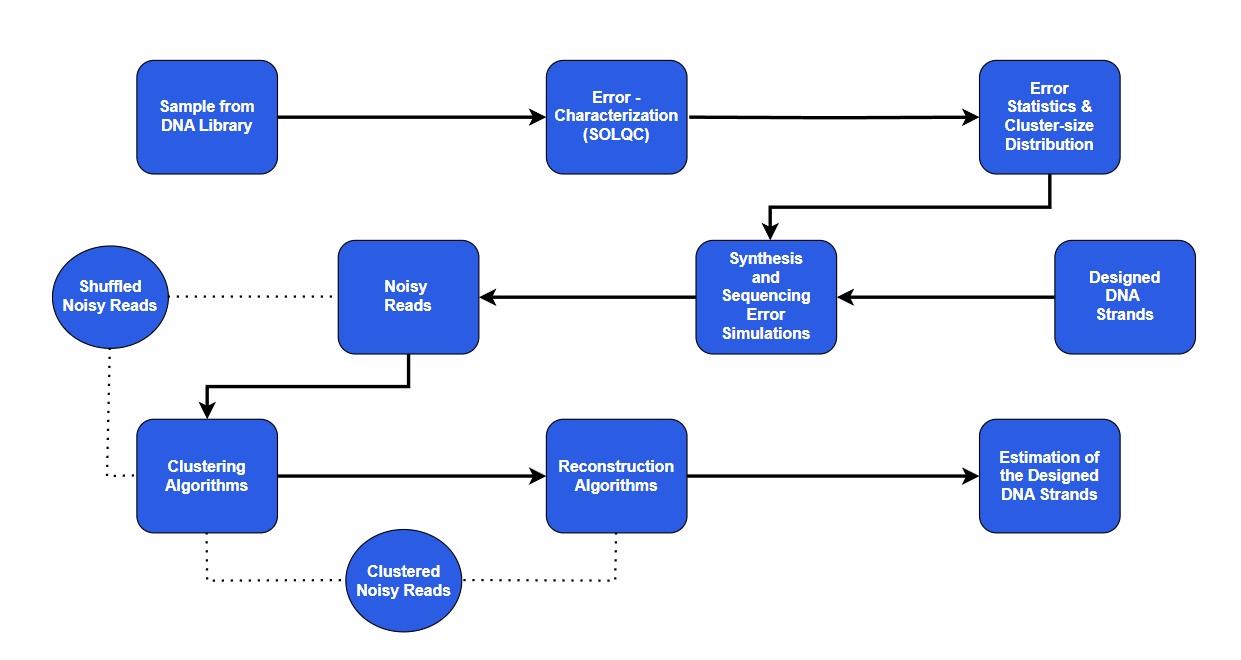}
\caption{Complete workflow of Storalator software. It is divided into four parts. 1. SOLQC or error characterization, 2. Error simulation, which is done by synthesis, PCR, and sequencing, 3. Clustering, and 4. Reconstruction. Image inspired from workshop on Non-Volatile memory by Omer Sabary, Gadi Chaykin, Nili Furman, Dvir Ben Shabat, and Eitan Yaakobi, 2022 \cite{storalatorpaper}. © Authors, Reprinted with permission.}
\label{storalatorRoadMap}
\end{figure*}

Storalator \cite{chaykin2022dna} accepts .txt file (containing only quaternary symbols) as an input (but does not accept .fasta file) and produces .txt file (which contains quaternary code words) as an output. It simulates real-life flow from DNA synthesis, sequencing, clustering, and reconstruction of DNA strands (doesn't take into account: encoding, decoding and storage, see Fig.~({\ref{Simulatorsblockdiagram}}) for its high level block diagram). Manual errors can be introduced from error statistics by users as per their own needs. Fig.~({\ref{storalatorRoadMap}}) shows the whole road map of the software. It can primarily be divided into four parts: 1. SOLQC or error characterization, 2. Error simulation, which is done by synthesis, PCR, and sequencing, 3. Clustering, and 4. Reconstruction. Their explanations are provided in this subsection only.

\subsubsection{Algorithms}
\begin{figure*}[h]
	\centering
	\includegraphics[scale=0.7]{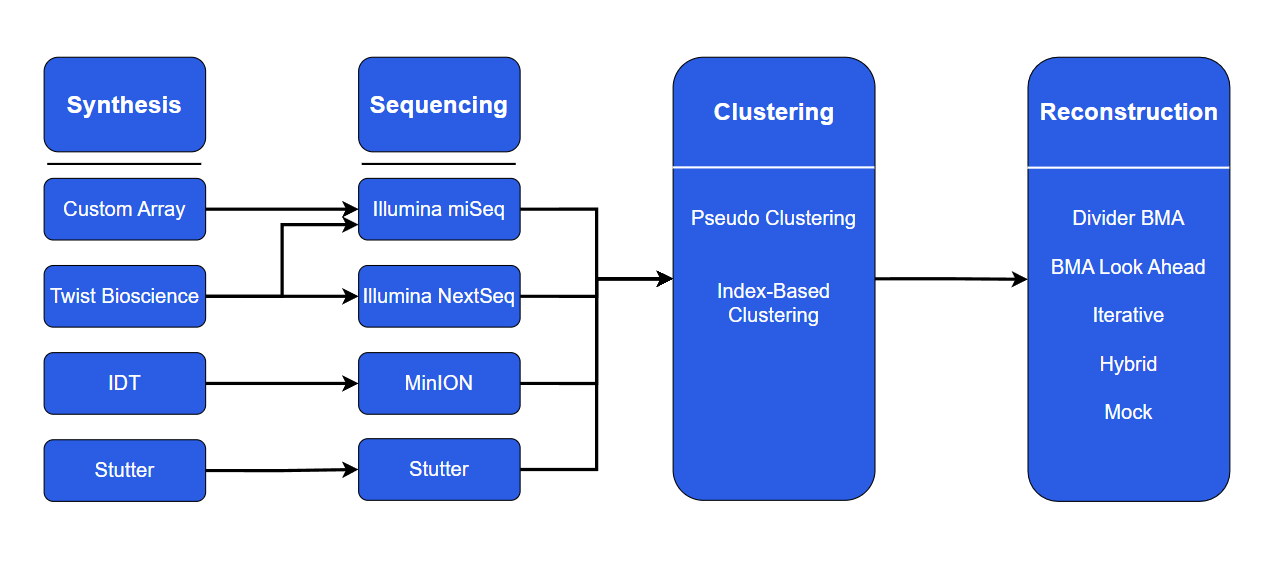}
\caption{Available algorithms in storalator software. Note that each synthesis technology can only be used with particular sequencing technology(s)}
\label{Storalatoralgorithms}
\end{figure*}

Fig.~({\ref{Storalatoralgorithms}}) shows all the algorithms available in storalator software. Each synthesis algorithm can only be used with particular sequencing algorithm(s).

\subsubsection{SOLQC}
It is a tool, which gives an analysis of the synthetic oligo libraries (Oligo libraries can be referred to as DNA barcodes). It gives out statistical information such as variance, error rates, etc. It can also give a graphical representation.

\begin{figure*}[h]
  \subcaptionbox*{Default interface}[.45\linewidth]{%
    \includegraphics[width=\linewidth]{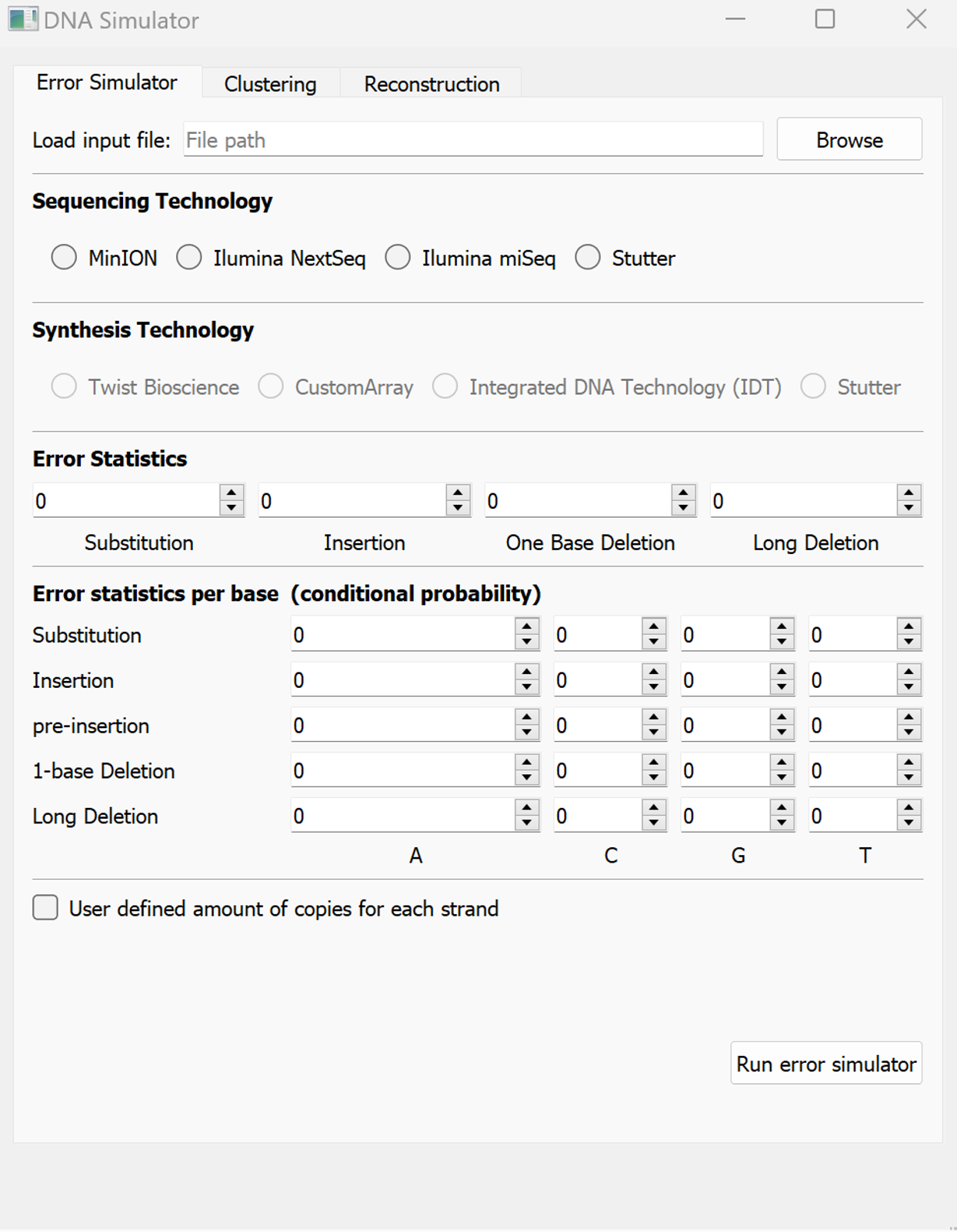}%
  }%
  \hfill
  \subcaptionbox*{Selecting Illumina NextSeq and Twist Bioscience, and generating error Statistics}[.45\linewidth]{%
    \includegraphics[width=\linewidth]{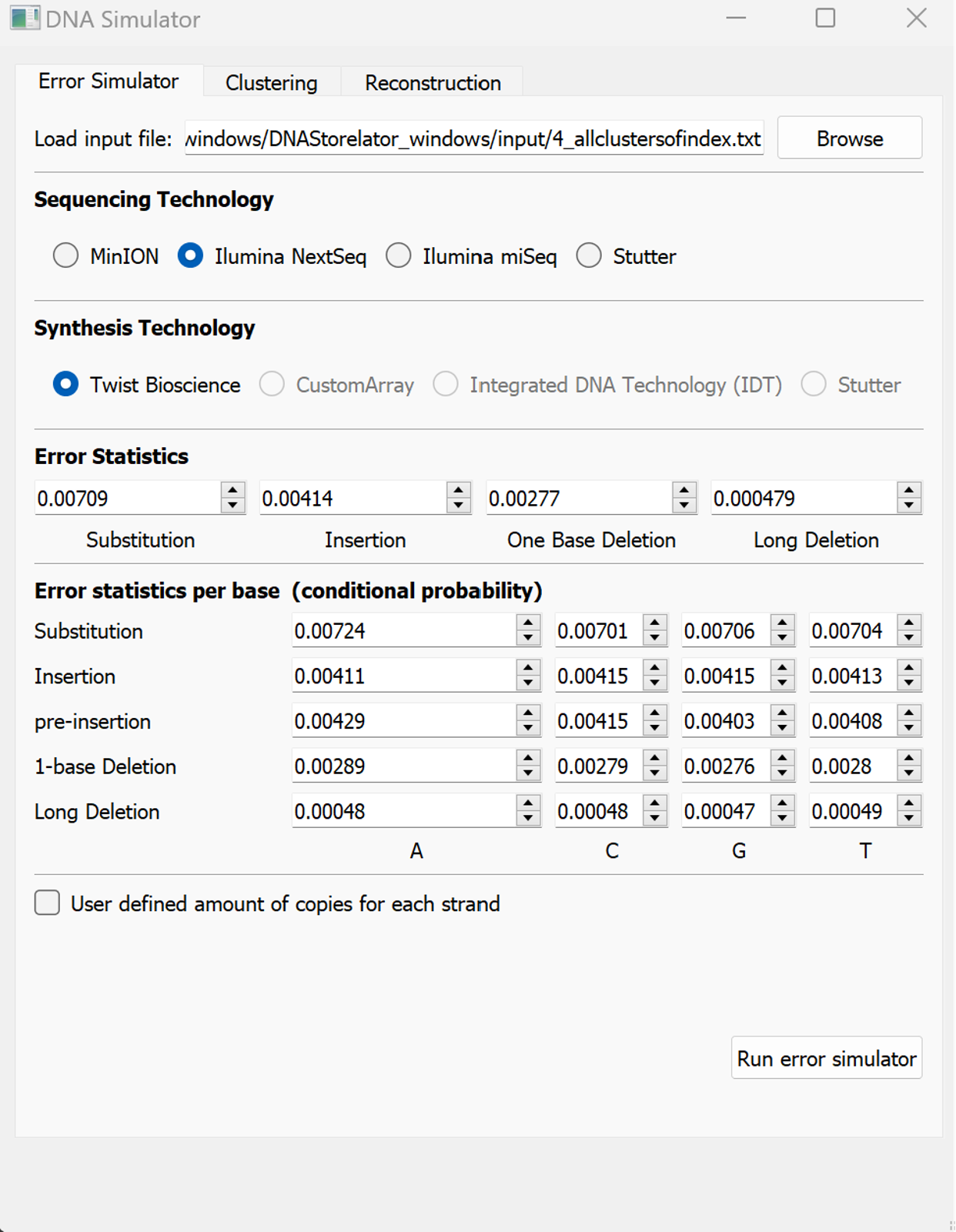}%
  }
  \caption{Storalator Error Simulator, screenshot taken from Storalator software developed by Omer Sabari, Eitan Yaakobi, Gadi Chaykin, Nili Furman \cite{storalatormainsoftware} © Authors, Reprinted with permission.}
  \label{Storalator_error_sim}
\end{figure*}

\begin{figure*}[h]
  \subcaptionbox*{Default clustering phase}[.32\linewidth]{%
    \includegraphics[width=\linewidth]{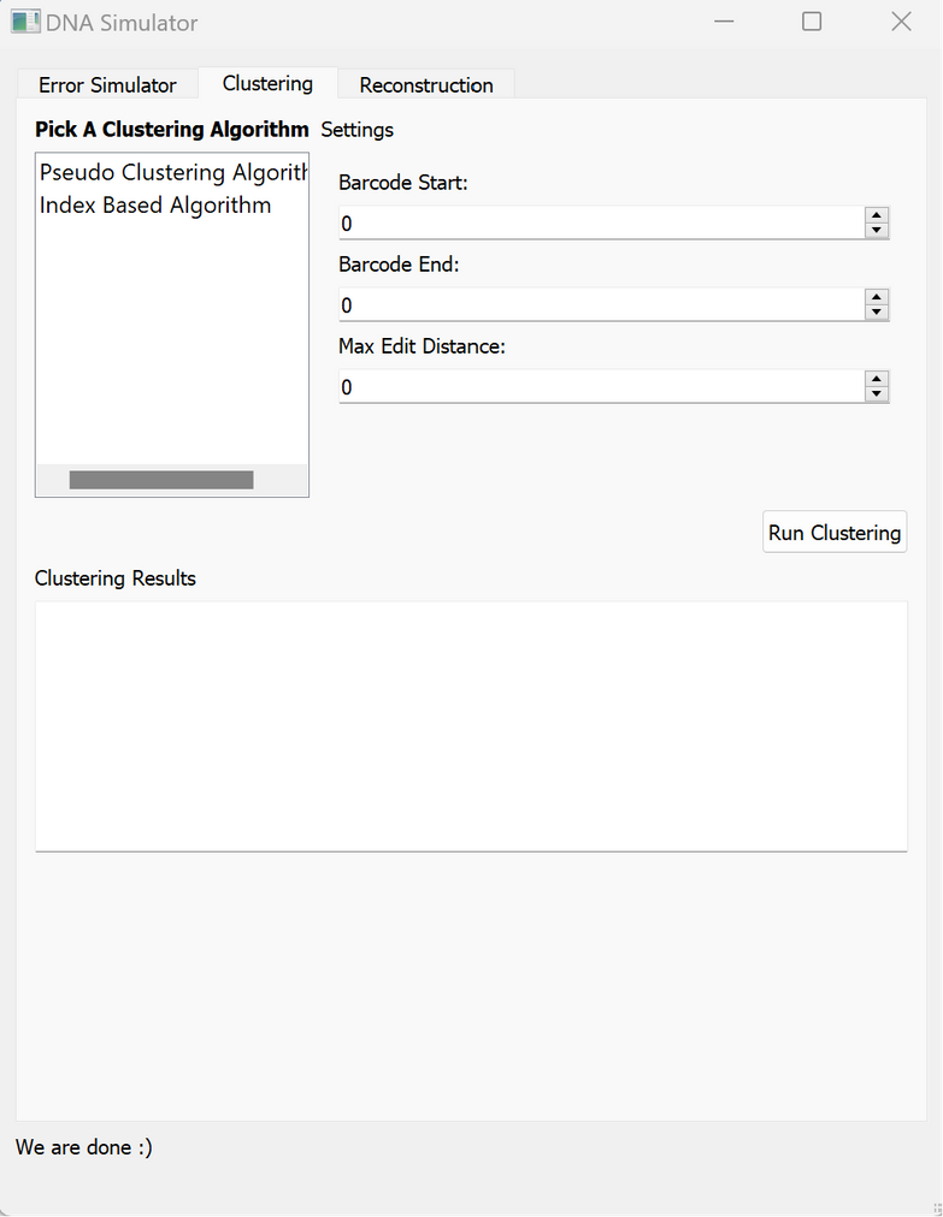}%
  }%
  \hfill
  \subcaptionbox*{Selecting index-based algorithm for clustering, and selecting the technologies we used before, along with the clustering index size}[.32\linewidth]{%
    \includegraphics[width=\linewidth]{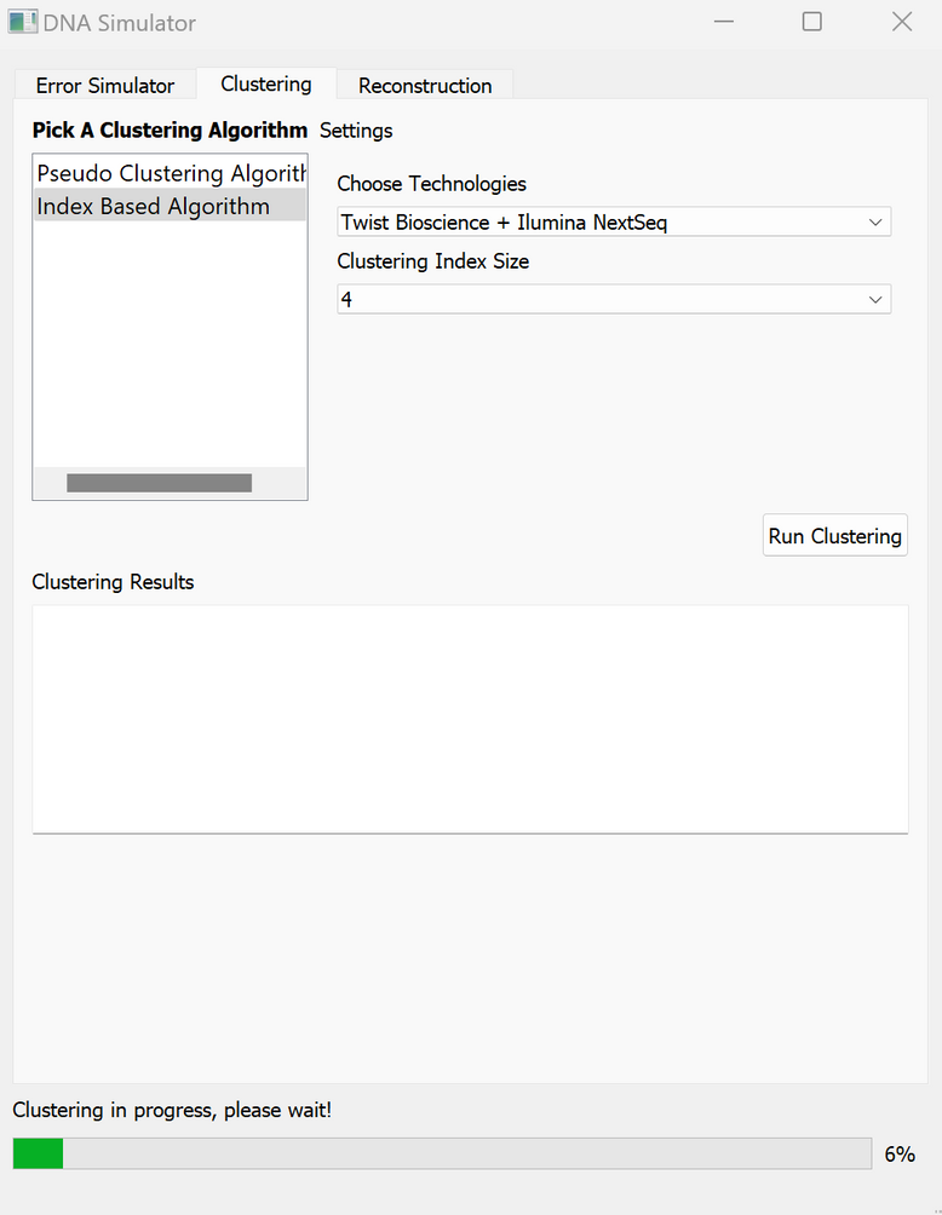}%
  }%
  \hfill
  \subcaptionbox*{Clustering results}[.32\linewidth]{%
    \includegraphics[width=\linewidth]{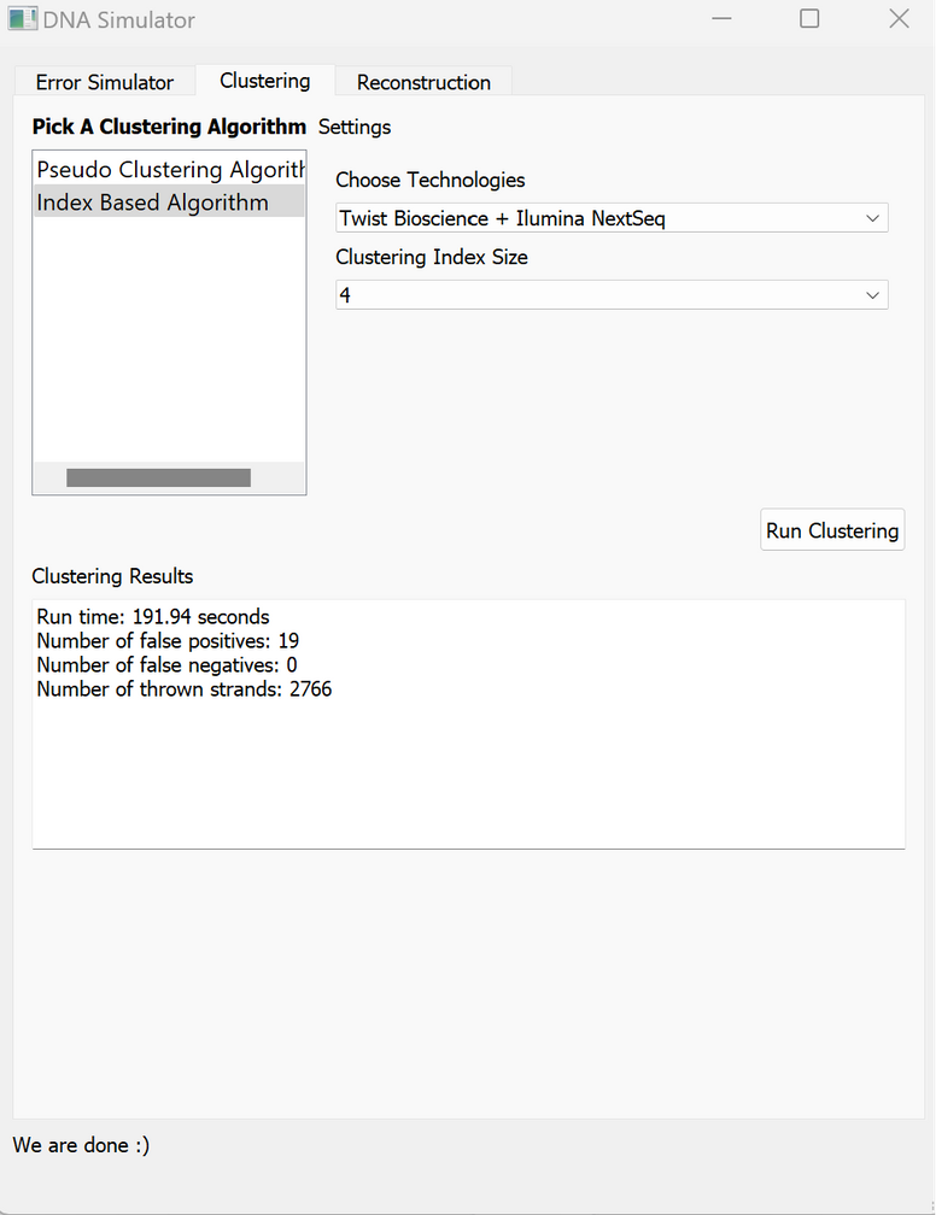}%
    }
  \caption{Storalator Clustering, screenshot taken from Storalator software developed by Omer Sabari, Eitan Yaakobi, Gadi Chaykin, Nili Furman \cite{storalatormainsoftware}. © Authors, Reprinted with permission.}
  \label{Storalator_clustering}
\end{figure*}

\begin{figure*}[h]
  \subcaptionbox*{Reconstruction errors histogram for BMA Look ahead}[.32\linewidth]{%
    \includegraphics[width=\linewidth]{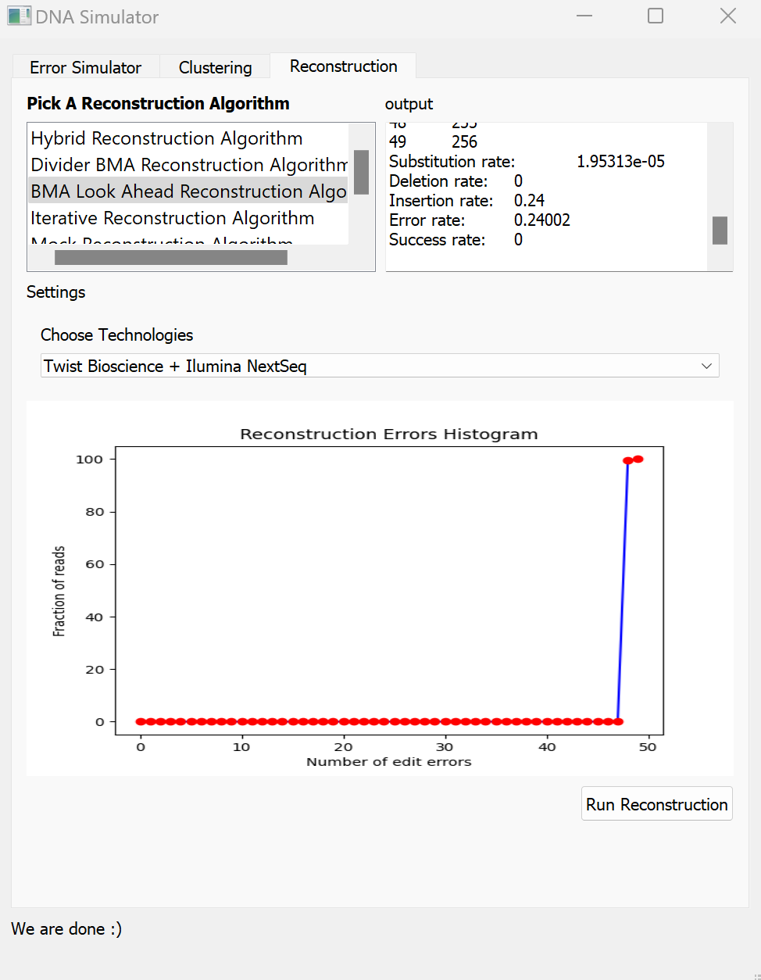}%
  }%
  \hfill
  \subcaptionbox*{Reconstruction errors histogram for Hybrid}[.32\linewidth]{%
    \includegraphics[width=\linewidth]{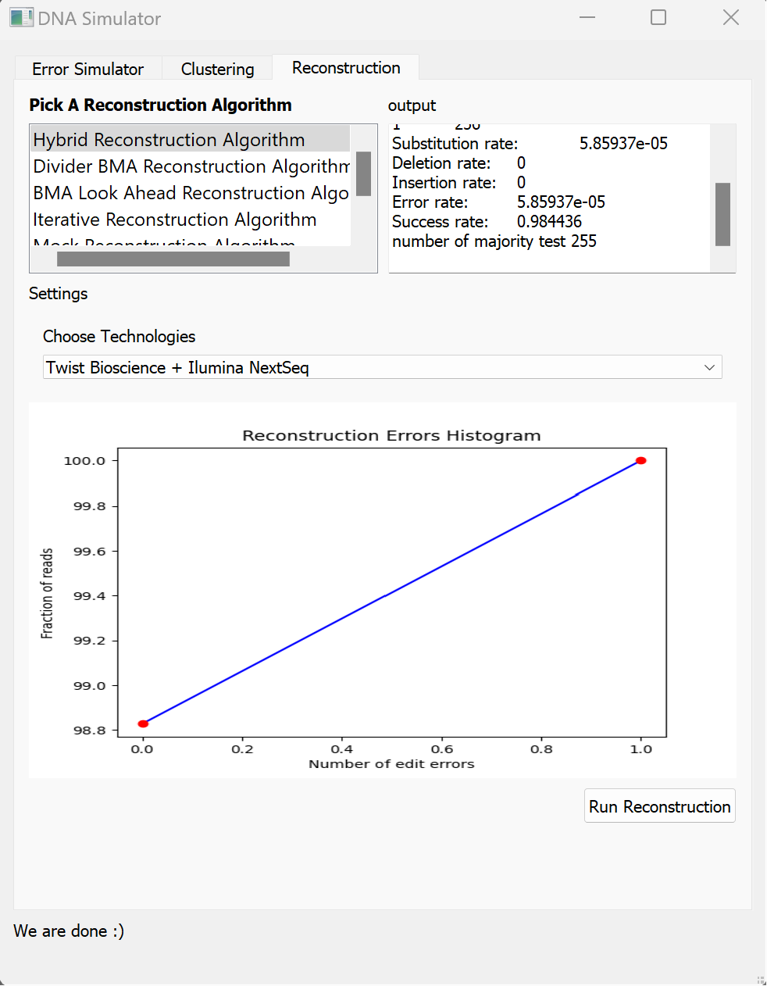}%
  }%
  \hfill
  \subcaptionbox*{Reconstruction errors histogram for Divider BMA}[.32\linewidth]{%
    \includegraphics[width=\linewidth]{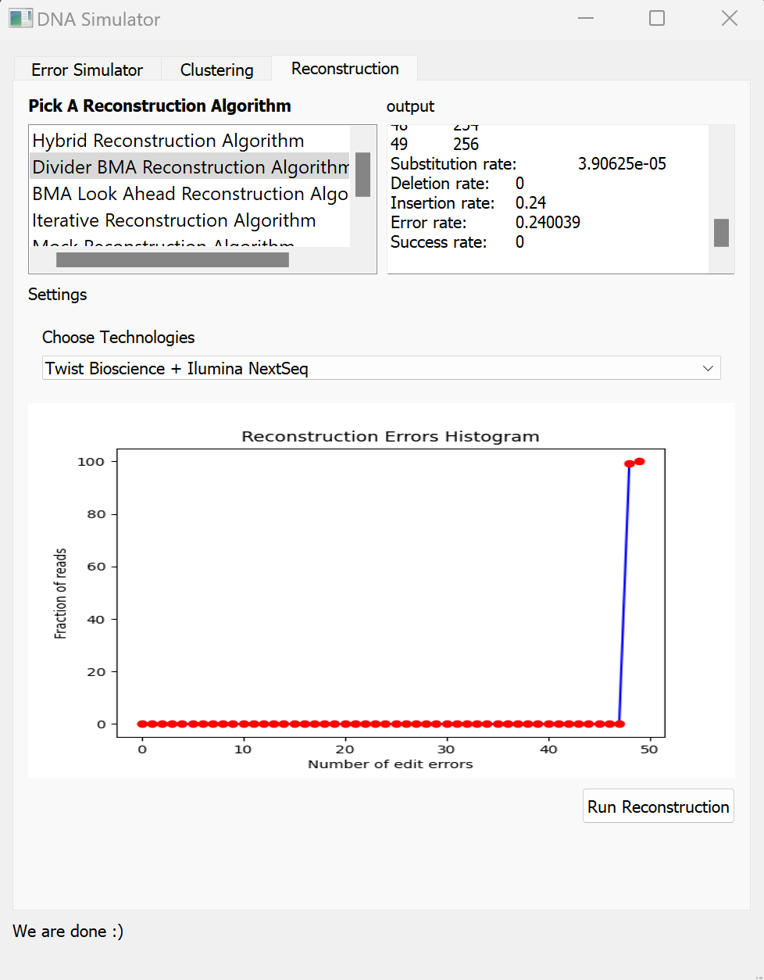}%
    }
  \caption{Selection of Reconstruction Algorithm for Illumina NextSeq and TwistBioscience, screenshot taken from Storalator software developed by Omer Sabari, Eitan Yaakobi, Gadi Chaykin, Nili Furman \cite{storalatormainsoftware}. © Authors, Reprinted with permission.}
  \label{Storalator_reconst_IlluminaNextSeq_TwistBioscience}
\end{figure*}

\begin{figure*}[h]
  \subcaptionbox*{Reconstruction errors histogram for BMA Look ahead}[.32\linewidth]{%
    \includegraphics[width=\linewidth]{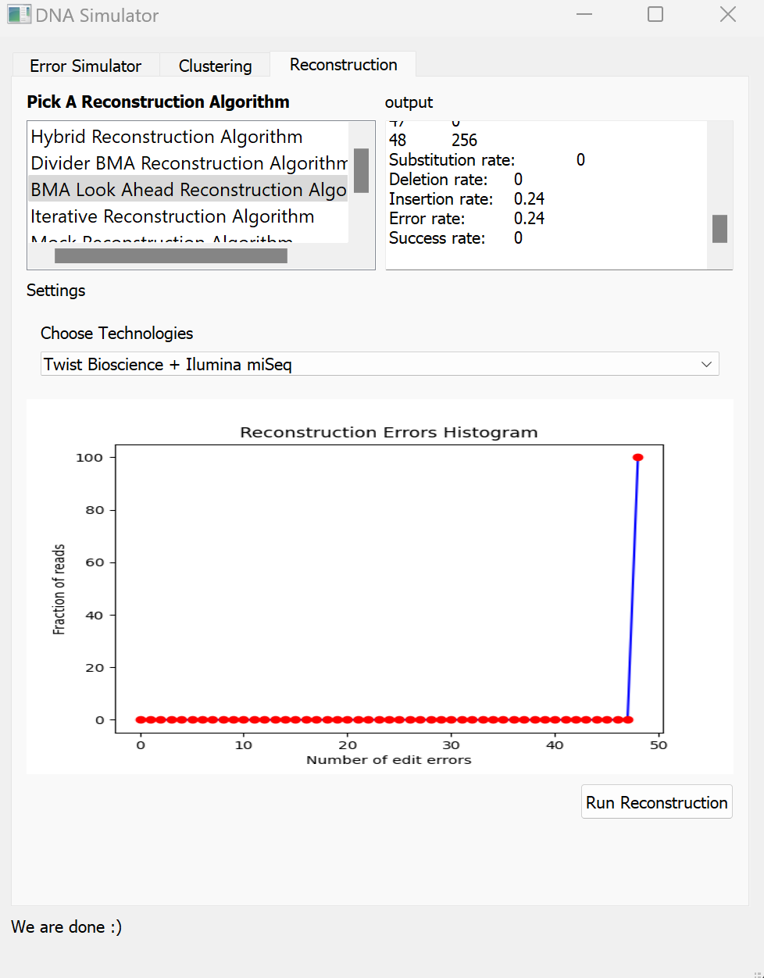}%
  }%
  \hfill
  \subcaptionbox*{Reconstruction errors histogram for Hybrid}[.32\linewidth]{%
    \includegraphics[width=\linewidth]{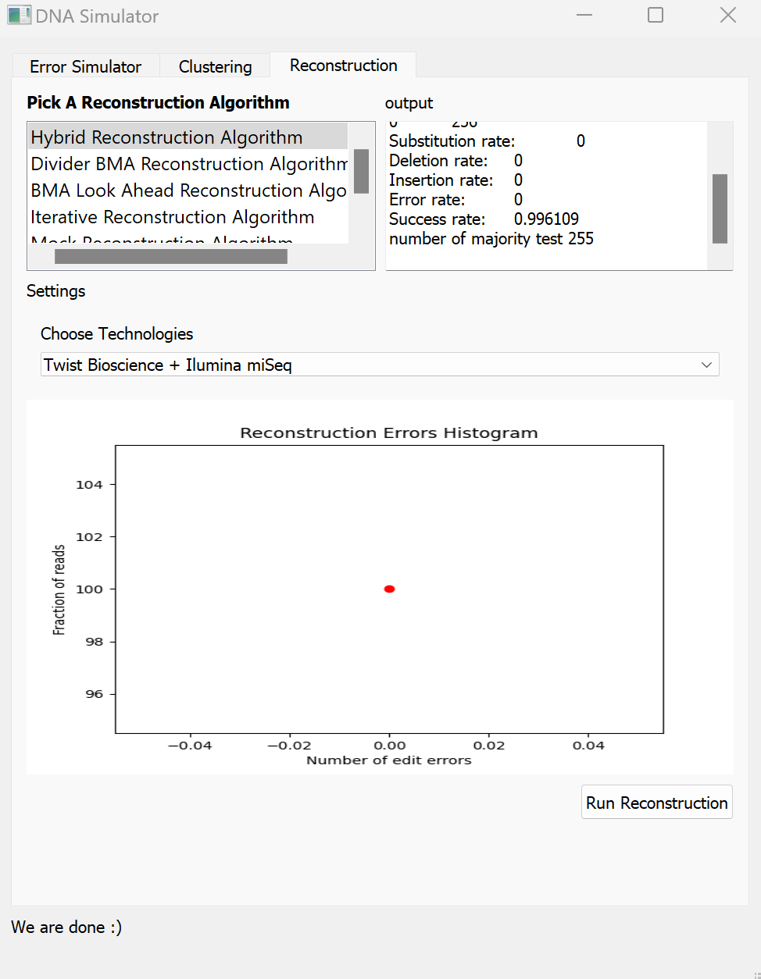}%
  }%
  \hfill
  \subcaptionbox*{Reconstruction errors histogram for Divider BMA}[.32\linewidth]{%
    \includegraphics[width=\linewidth]{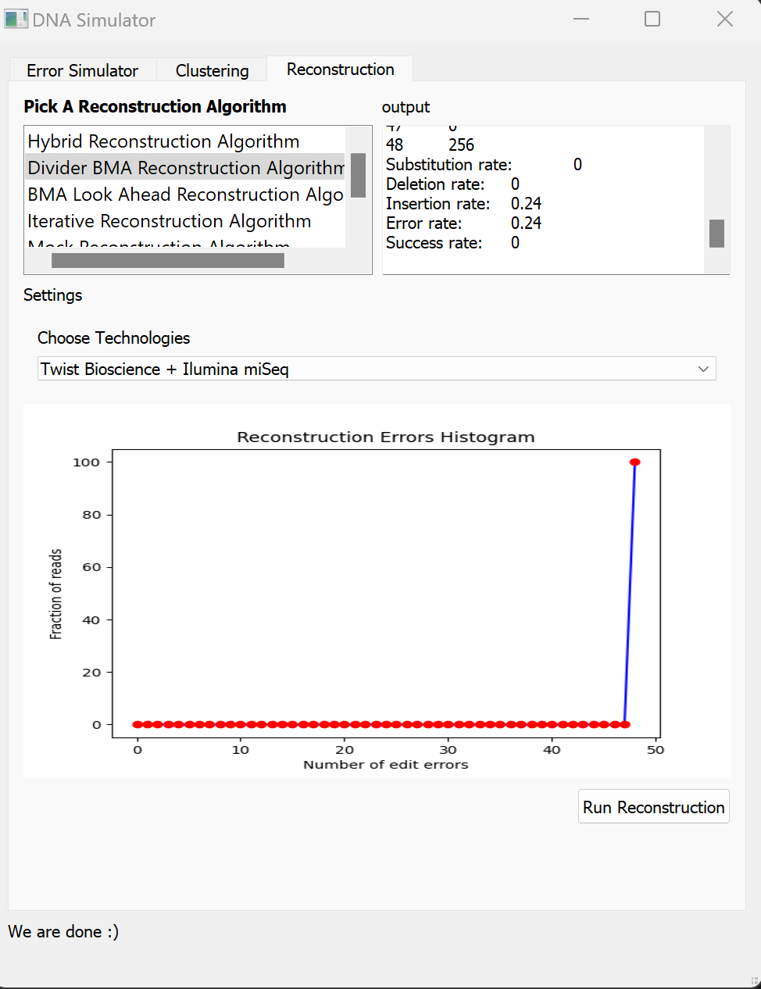}%
    }
  \caption{Selection of Reconstruction Algorithm for Illumina miSeq and TwistBioscience, screenshot taken from Storalator software developed by Omer Sabari, Eitan Yaakobi, Gadi Chaykin, Nili Furman \cite{storalatormainsoftware}. © Authors, Reprinted with permission.}
  \label{Storalator_reconst_IlluminamiSeq_TwistBioscience}
\end{figure*}

\begin{figure*}[h]
  \subcaptionbox*{Reconstruction errors histogram for BMA Look ahead}[.32\linewidth]{%
    \includegraphics[width=\linewidth]{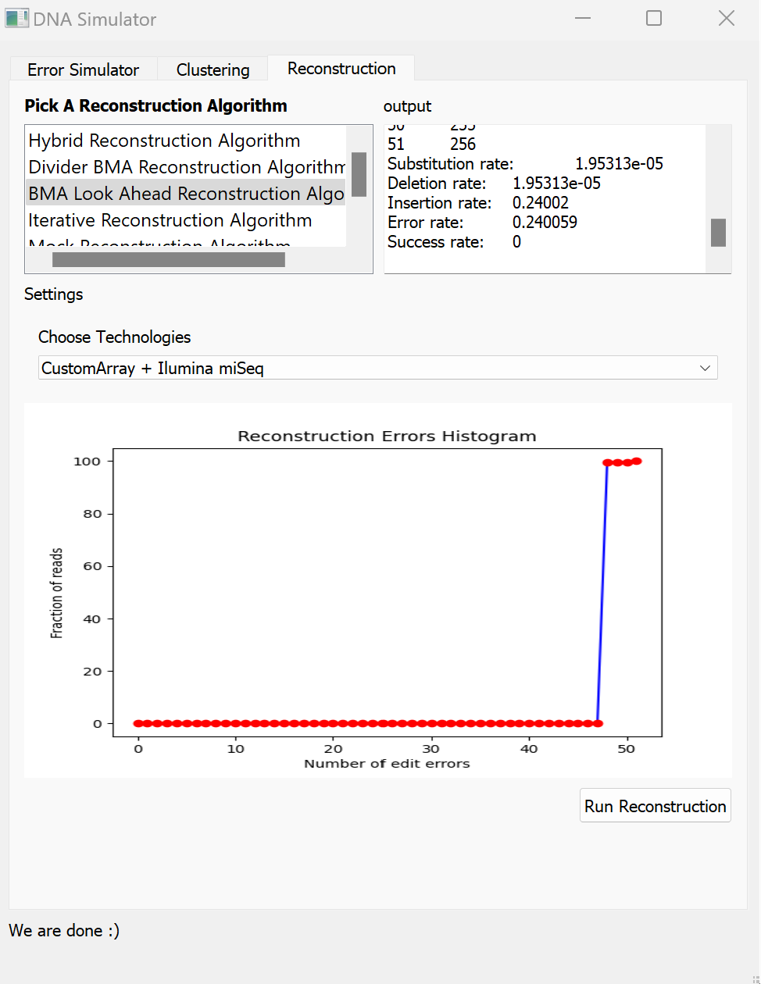}%
  }%
  \hfill
  \subcaptionbox*{Reconstruction errors histogram for Hybrid}[.32\linewidth]{%
    \includegraphics[width=\linewidth]{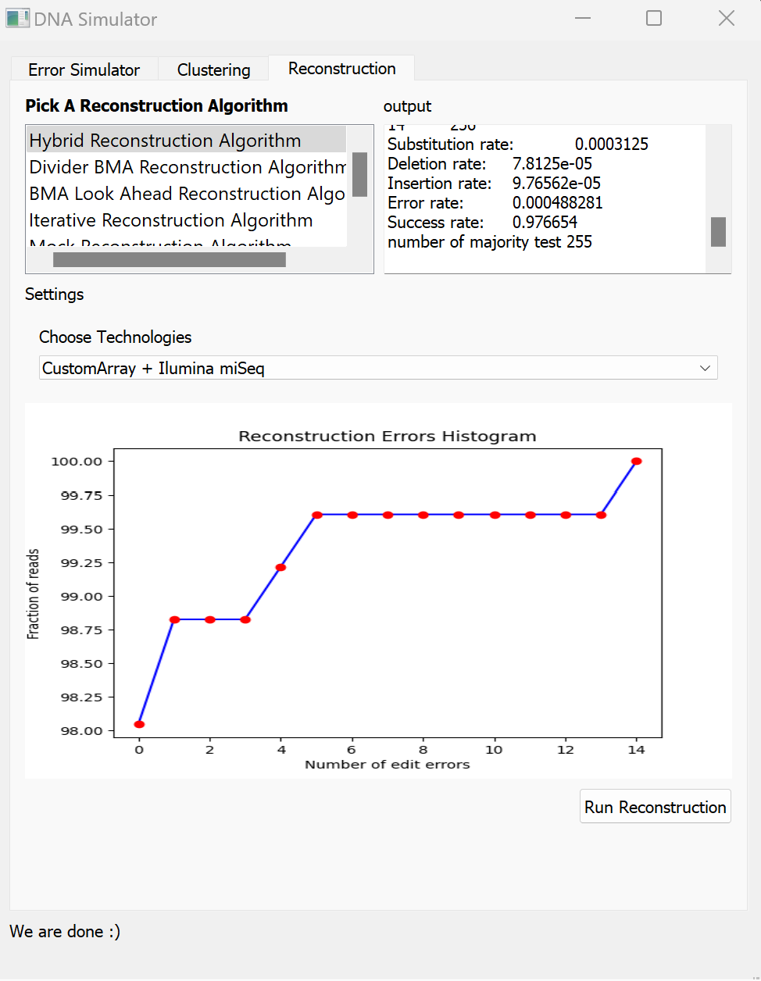}%
  }%
  \hfill
  \subcaptionbox*{Reconstruction errors histogram for Divider BMA}[.32\linewidth]{%
    \includegraphics[width=\linewidth]{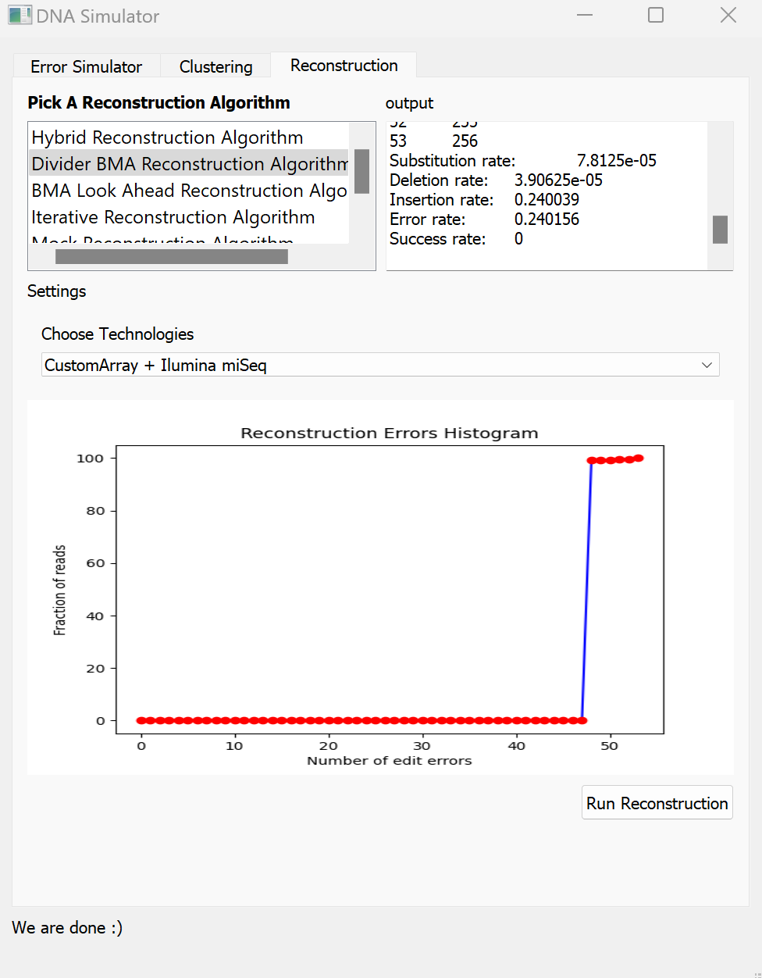}%
    }
  \caption{Selection of Reconstruction Algorithm for Illumina miSeq and CustomArray, screenshot taken from Storalator software developed by Omer Sabari, Eitan Yaakobi, Gadi Chaykin, Nili Furman \cite{storalatormainsoftware}. © Authors, Reprinted with permission.}
  \label{Storalator_reconst_IlluminamiSeq_CustomArray}
\end{figure*}

\begin{figure*}
	\centering
	\includegraphics[width=\linewidth]{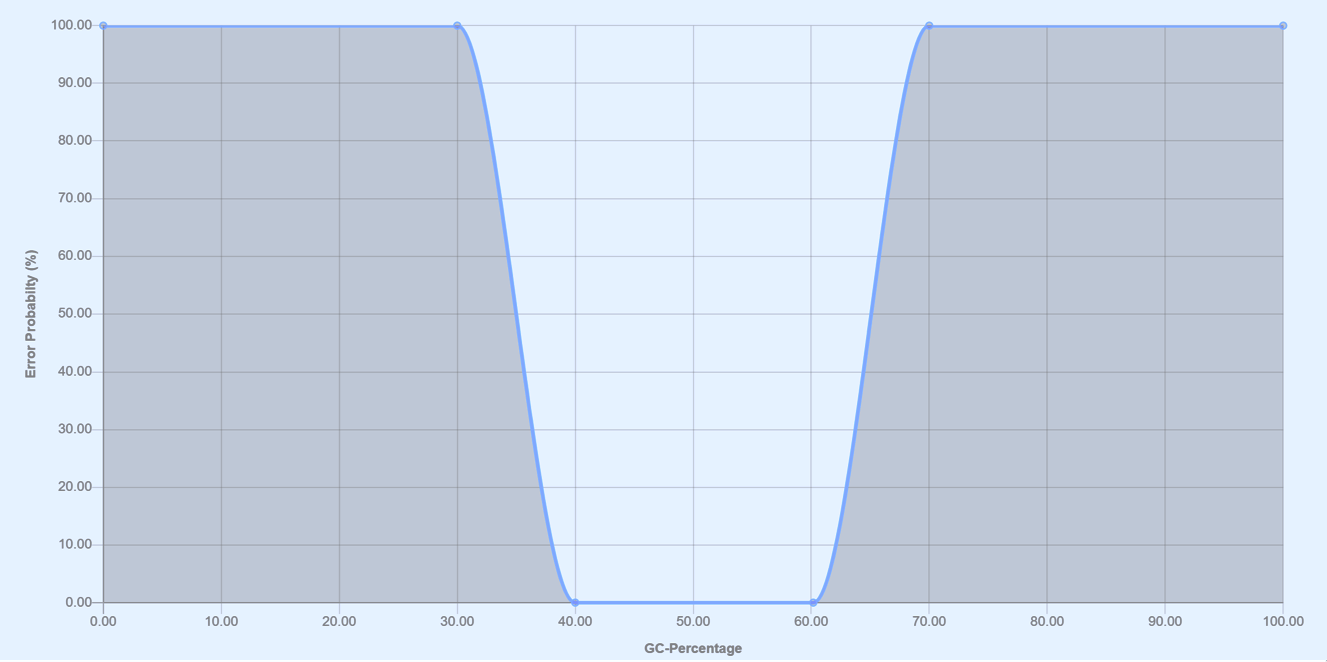}
\caption{It represents that when GC content is in its optimal range their will be less error probability, but when it is less/more than its optimal range error probability will spike up due to less/more hydrogen bonding resulting into error. Error probability vs GC-percentage, screenshot taken from MESA software developed by Schwarz et al. (image further enhanced for better visibility) \cite{mesamainsoftware}. © Authors, Reprinted with permission.}
\label{MESA_GC_graph}
\end{figure*}

\begin{figure*}[h]
  \subcaptionbox*{Reconstruction errors histogram for BMA Look ahead}[.32\linewidth]{%
    \includegraphics[width=\linewidth]{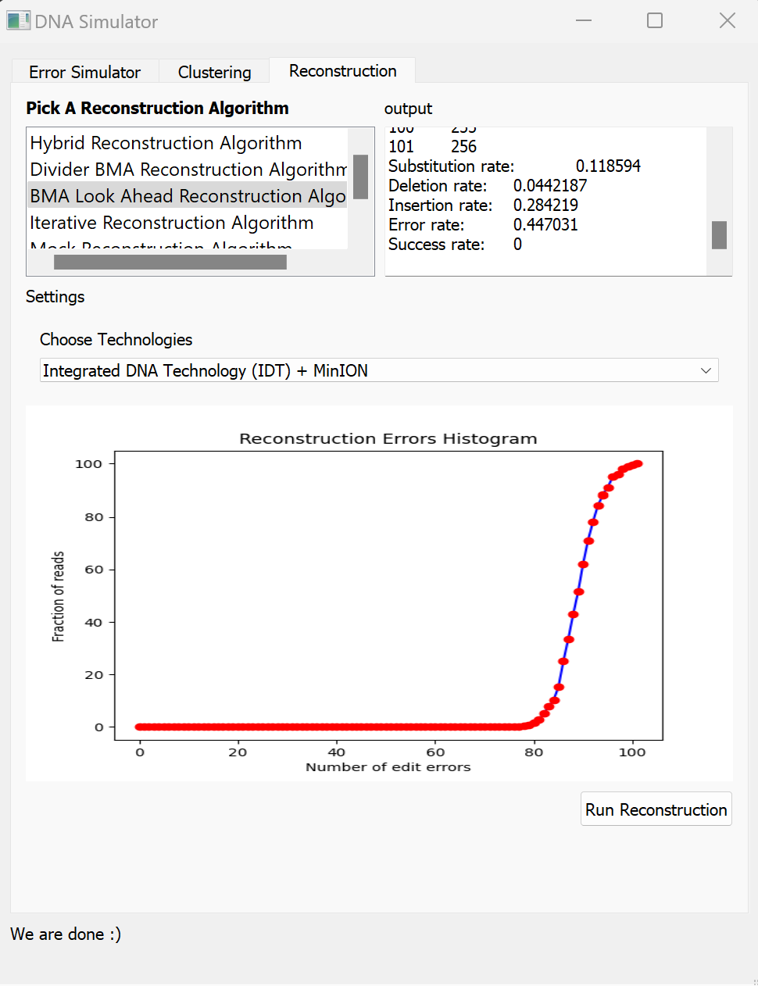}%
  }%
  \hfill
  \subcaptionbox*{Reconstruction errors histogram for Hybrid}[.32\linewidth]{%
    \includegraphics[width=\linewidth]{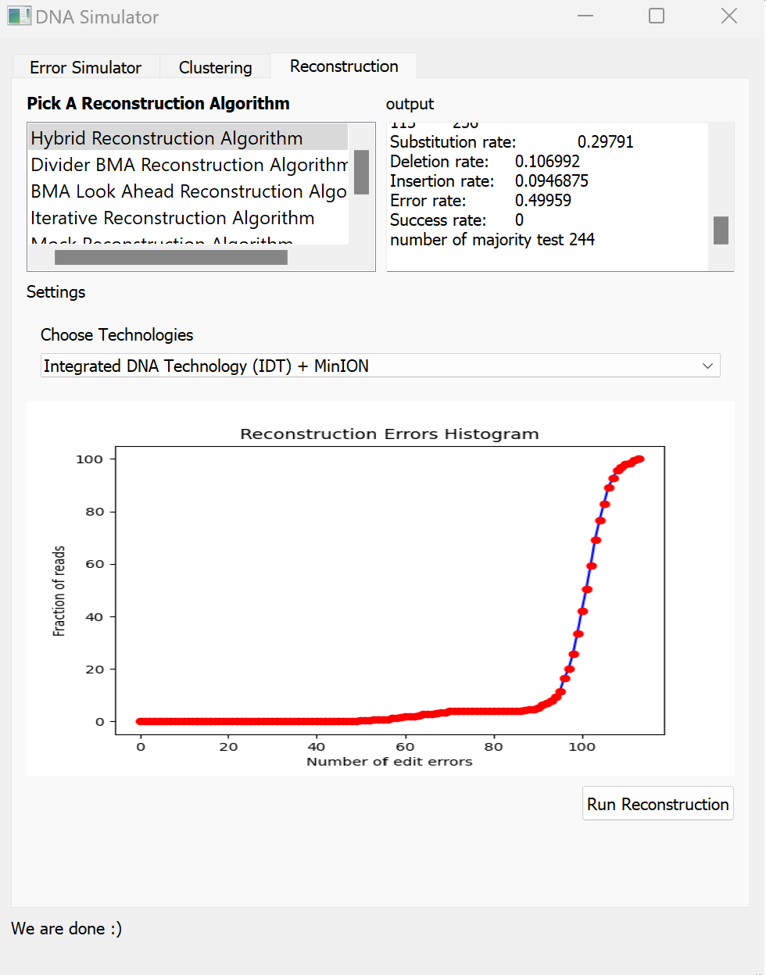}%
  }%
  \hfill
  \subcaptionbox*{Reconstruction errors histogram for Divider BMA}[.32\linewidth]{%
    \includegraphics[width=\linewidth]{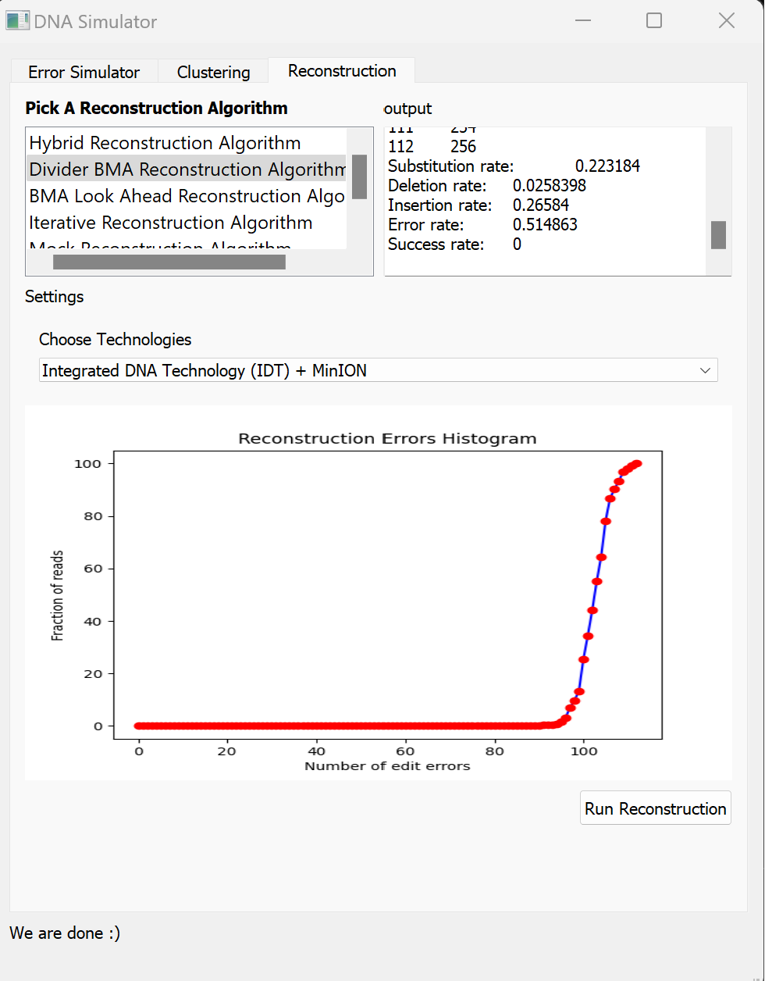}%
    }
  \caption{Selection of Reconstruction Algorithm for MinION and IDT, screenshot taken from Storalator software developed by Omer Sabari, Eitan Yaakobi, Gadi Chaykin, Nili Furman \cite{storalatormainsoftware}. © Authors, Reprinted with permission.}
  \label{Storalator_reconst_MinION_IDT}
\end{figure*}

\begin{figure*}[h]
  \subcaptionbox*{Reconstruction errors histogram for BMA Look ahead}[.32\linewidth]{%
    \includegraphics[width=\linewidth]{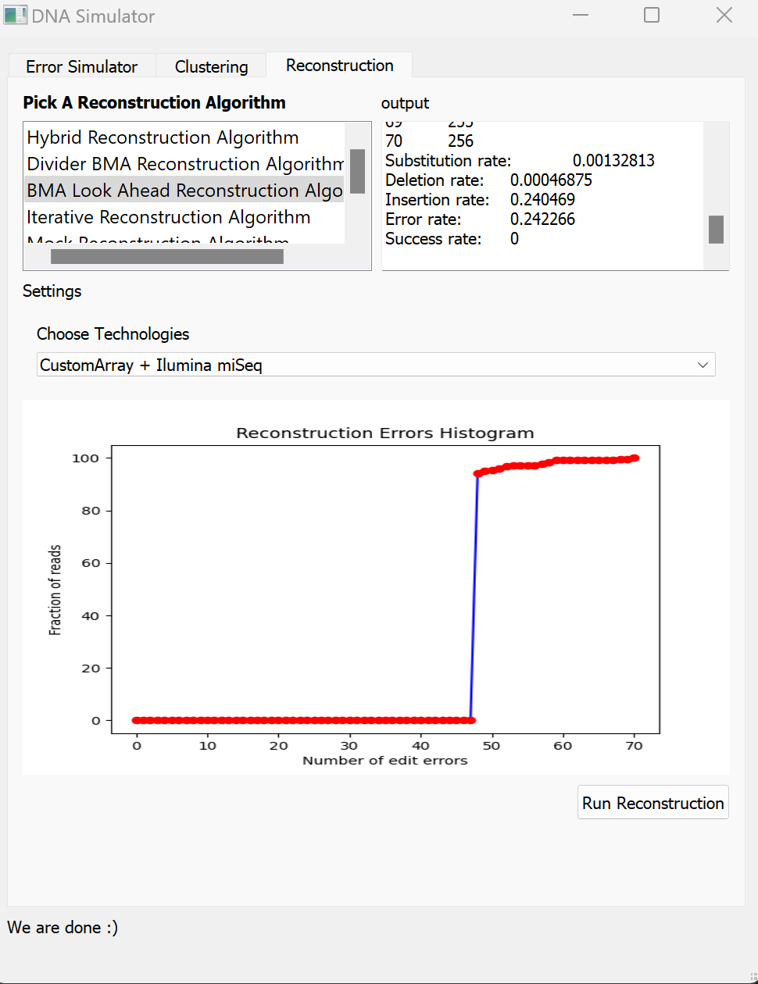}%
  }%
  \hfill
  \subcaptionbox*{Reconstruction errors histogram for Hybrid}[.32\linewidth]{%
    \includegraphics[width=\linewidth]{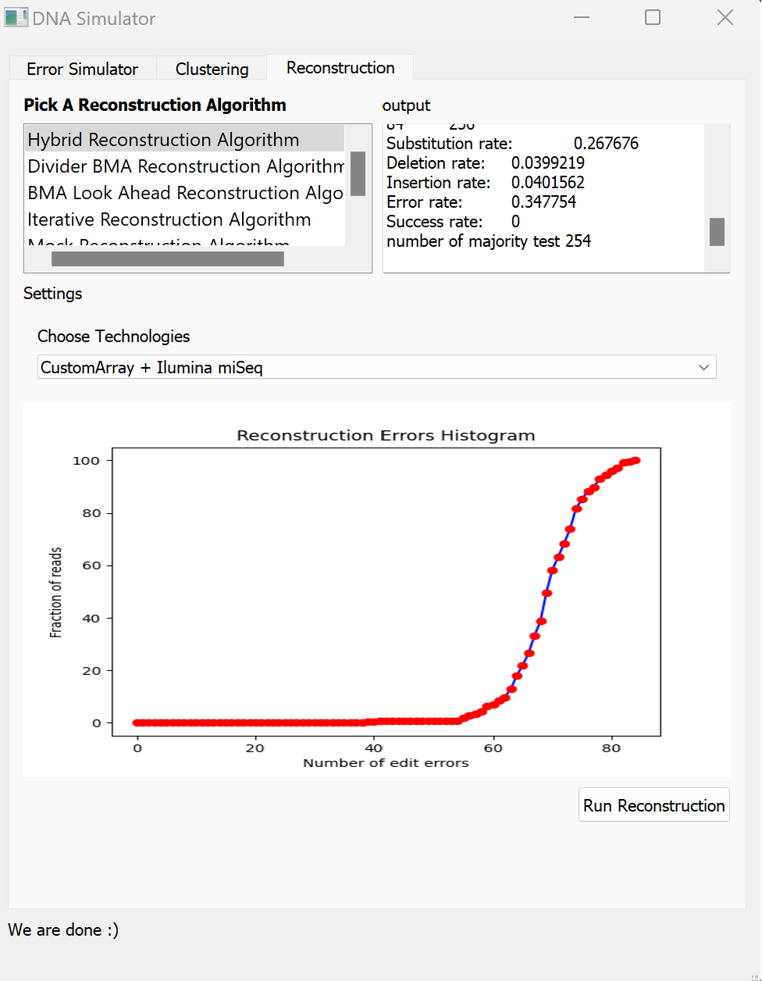}%
  }%
  \hfill
  \subcaptionbox*{Reconstruction errors histogram for Divider BMA}[.32\linewidth]{%
    \includegraphics[width=\linewidth]{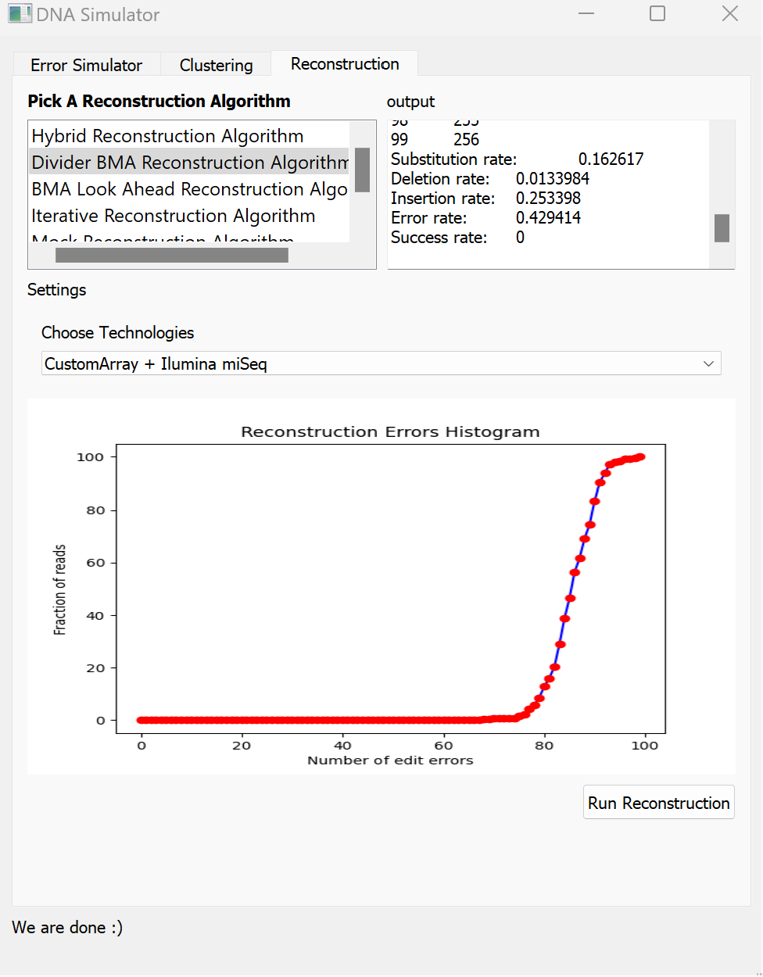}%
    }
  \caption{Selection of Reconstruction Algorithm for Stutter and Stutter, screenshot taken from Storalator software developed by Omer Sabari, Eitan Yaakobi, Gadi Chaykin, Nili Furman \cite{storalatormainsoftware}. © Authors, Reprinted with permission.}
  \label{Storalator_reconst_Stutter_Stutter}
\end{figure*}

\begin{figure*}
	\centering
	\includegraphics[scale=0.80]{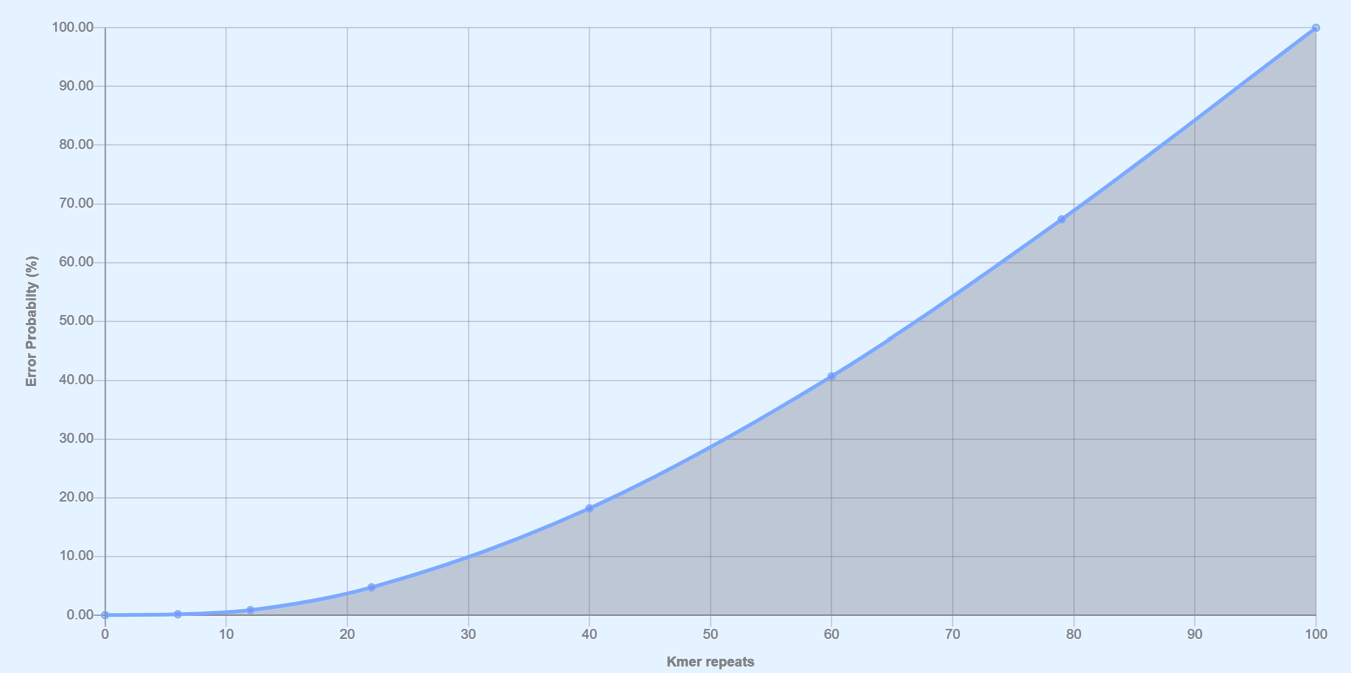}
\caption{It represents that when K-mer sequence repetition is more then the error probability gradually increases during sequencing. Error probability vs Kmer-repeats, screenshot taken from MESA software developed by Schwarz et al. (image further enhanced for better visibility) \cite{mesamainsoftware}. © Authors, Reprinted with permission.}
\label{MESA_Kmer_graph}
\end{figure*}

\begin{figure*}
	\centering
	\includegraphics[scale=0.79]{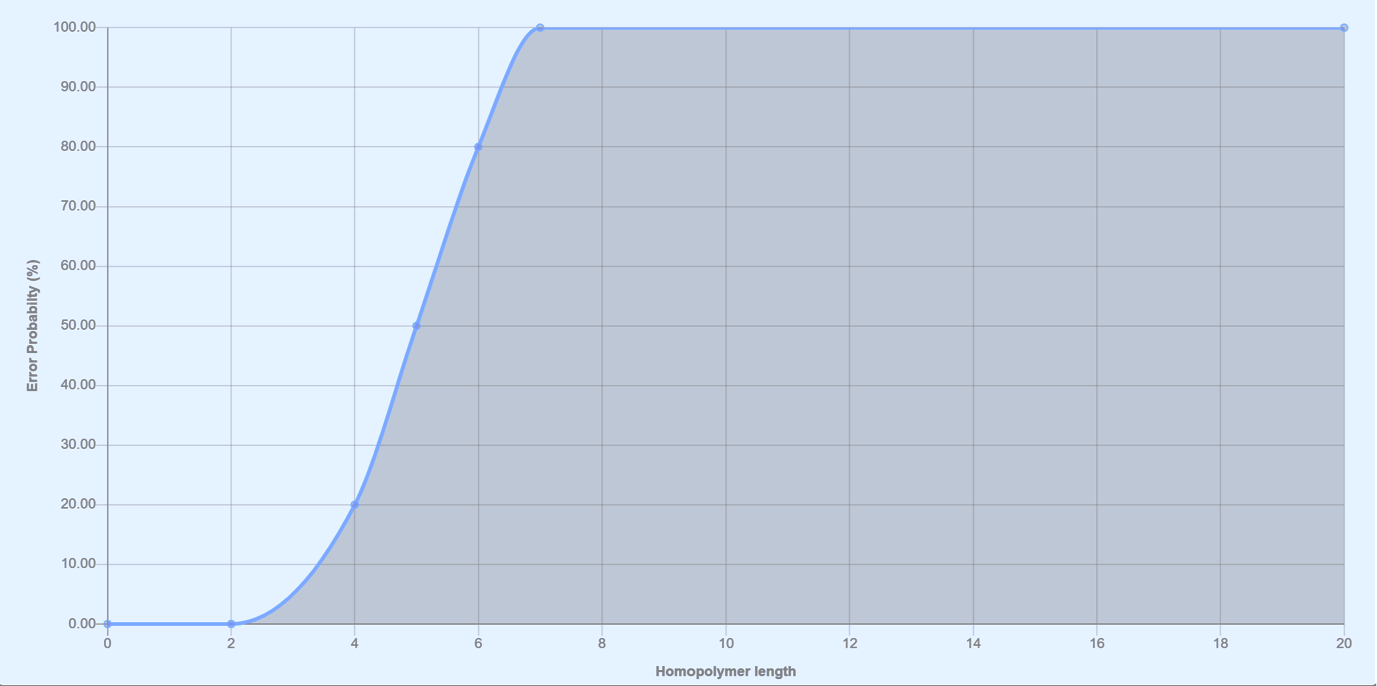}
\caption{It represents that as the length of the Homopolymer increases error probability increases rapidly, as Homopolymers causes a lot of hinderance during reading and thus results into IDS errors. Error probability vs Homopolymer length, screenshot taken from MESA software developed by Schwarz et al. (image further enhanced for better visibility) \cite{mesamainsoftware}. © Authors, Reprinted with permission.}
\label{MESA_homopolymer_graph}
\end{figure*}

\subsubsection{Synthesis}
\textbf{Twist Bioscience} \cite{twistbioscience} is a company that synthesizes DNA on silicon instead of doing it traditionally on 96-well plastic plates. It can produce 1-100,000 genes in parallel, in the same amount of time. Its cost can go as low as \$0.07 per bp, and \$0.09 per bp for Next Generation Sequencing. It uses phosphoramidite chemistry to synthesize large amounts of oligonucleotide sequences for up to 200 bp in length.

\textbf{CustomArray} \cite{customarray} is a synthetic biology company that uses semiconductor chip technology to synthesize DNA electrochemically on the chip surface. The B3TM Synthesizer uses semiconductor arrays to synthesize DNA arrays on the chip.

\textbf{Integrated DNA Technologies (IDT)} \cite{idt} synthesizes DNA and RNA oligos using a method called phosphoramidite addition. It uses a controlled-pore glass (CPG) solid support and amide chemistry to synthesize oligos in the 3’ $\rightarrow$ 5’ direction.

\textbf{Stutter} \cite{stutter} is a byproduct of amplifying short tandem repeats (STR). DNA sequences of two to six base pairs in length are known as STRs. They are referred to as simple sequence repeats or microsatellites. Stutter occurs when a minor product is generated that is typically one repeat smaller than the primary allele. Stutter peaks are small and appear right before and after the real peak of PCR. They occur when the polymerase slips forward or backward four base pairs during the copying of a DNA strand. For further information on enzymatic synthesis interested readers may refer to \cite{enzymaticsynthesis} \cite{terminatorfreeenzymaticDNAsynthesis}.

\subsubsection{Sequencing}

\textbf{Illumina MiSeq} \cite{illuminamiseq} is a next-generation sequencing instrument that uses sequencing-by-synthesis technology, which detects single bases by a fluorescent label as they are incorporated into the growing DNA strands.

\textbf{Illumina NextSeq} \cite{illuminanextseq} is a sequencing system that includes: Exome sequencing, Target enrichment, Single-cell profiling, and transcriptome sequencing. NextSeq 1000 and NextSeq 2000 use patterned flow cells. NextSeq 500 has the accuracy of Illumina SBS technology. With its reliable RNA-to-results approach, the NextSeq 550 system offers RNA sequencing applications ranging from whole-transcriptome analysis to gene expression profiling.

\textbf{MinION} \cite{minion} is a portable, real-time DNA-RNA sequencing device. Its size is smaller than a phone. We just need to add a sample to the flow cell. Then the pore signal is analyzed as the molecule will be analyzed. Also, it’s very easy to use as it can directly connect to a laptop. It weighs less than 100g, has 512 Nano pore channels, and can achieve 30 gigabases of sequence data per flow cell.

\textbf{Stutter} \cite{stutter} uses two sequencing algorithms: HipSTR \cite{willems2017genome} \& STRsearch \cite{wang2020strsearch}. HipSTR uses an alignment model which accounts for Illumina sequencing errors and STR-specific errors. STRsearch uses a strategy of counting repeat patterns (motifs) and INDELs in the repeat region.

\subsubsection{Reconstruction}
Three main reconstruction algorithms used in storalator: Iterative Reconstruction, Divider BMA, BMA Lookahead. Interested readers may refer to the document \cite{qin2022robust} \cite{tracereconstruction}.




\subsubsection{Simulations}
Simulations occur in three phases: error simulation, clustering, and reconstruction. We performed simulations for all three phases and some of them are given below with images. Note that these simulations were performed on a device having 16GB of RAM, and an i7-11th gen intel processor.

\begin{enumerate}
    \item Error simulation: Error statistics for Illumina NextSeq and Twist Bioscience selection (see Fig.~({\ref{Storalator_error_sim}}))
    \item Clustering: Shows the result for index-based clustering algorithm along with Illumina NextSeq and Twist Bioscience (see Fig.~{(\ref{Storalator_clustering}}))
    \item Reconstruction : Simulation results are shown for 5 synthesis-sequencing pairs and for three reconstruction algorithms: BMA Look Ahead, Hybrid, and Divider BMA. Illumina NextSeq and TwistBioscience (see Fig. ~({\ref{Storalator_reconst_IlluminaNextSeq_TwistBioscience}}), Illumina miSeq and TwistBioscience (see Fig. ~({\ref{Storalator_reconst_IlluminamiSeq_TwistBioscience}}), Illumina miSeq and CustomArray (see. Fig. ~({\ref{Storalator_reconst_IlluminamiSeq_CustomArray}}), MinION and IDT (see Fig.~({\ref{Storalator_reconst_MinION_IDT}}), and Stutter and Stutter (see Fig.~({\ref{Storalator_reconst_Stutter_Stutter}})
\end{enumerate}

\subsubsection{Improvements}
Storalator only accepts ACGT-format files as an input and produces ACGT string as an output. So there’s scope for including more algorithms for converting binary files to ACGT file format. Only .txt files are accepted by the software and no .fasta files, which are default file formats for DNA strings. There is no option for accessing the file after the insertion of errors, after clustering, or after reconstruction. Any custom files that already have some errors, or are already clustered for further processing cannot be given to this software. Further it does not take any kind of storage or temperature into account, for the storage of data into DNA.

\subsubsection{Speculated Errors}
Software crashes when a text file without ACGT is given. If there are 256 strands, 4-index clustering takes place. But this software gives options for 4-index, 5-index, and 9-index, wherein it works for 4-index but in other cases, it gives divide-by-zero errors thereby crashing the software. There are options for selecting technology (synthesis + sequencing) during the clustering and reconstruction stage, which are kind of not related to any clustering or reconstruction algorithm. There is no Stutter-Stutter synthesis and sequencing folder. In UI design, Sequencing is mentioned before Synthesis.

\subsection{MESA (Mosla Error Simulator)}
MESA \cite{schwarz2020mesa} is a simulator that includes synthesis methods, sequencing methods, the number of PCR cycles, which PCR to use, months of storage, and the storage host (Fig.~({\ref{Simulatorsblockdiagram}}) represent high level block diagram of MESA). It also gives a detailed analysis of errors, factors such as temperature, storage time, etc. It does not account for encoding, decoding, clustering, and reconstruction. Greater level of customization is present in MESA as compared to Storalator and DeepSimulator.

\subsubsection{GC and Kmer window}
GC window significantly represents the content of guanine and cytosine out of the total number of bases present in DNA. This technique can basically. count the number of GCs in comparison to ATs. Behaviour of PCR can be predicted by GC content. Kmer is a sequence of K nucleotides. It analyzes all the subsequences and checks if there are any repetitive subsequences present which helps in analyzing sequencing output. GC window and Kmer window are used for this software. In GC content, guanines tend to form hydrogen bonds. So, if the GC content increases, hydrogen bond tightens up and thus results in an increase in melting temperature. The increase in melting temperature is because of inter and intra-stranding. So, if GC content is out of range then the error probability increases. In the Kmer window, if formed K windows are repetitive, then the error probability increases. This happens as it causes hindrance in sequencing. Because of the high GC content, it results in a high melting point, which further affects the DNA fragment's separation during the PCR's denaturation phase. This will ultimately result in reduced yield for PCR, because it cannot effectively synthesize in the presence of hydrogen bonds. This will indeed result in polymerase slippage and will reconnect at a different position, thus contributing to secondary structures. Oxford Nanopore sequencing, PacBio sequencing, and Illumina sequencing technologies work best for GC content in optimal range. For out of range GC content, all of these technologies will show errors. Illumina sequencing and Oxford Nanopore are sensitive to homopolymers too. Common substitution patterns for PacBio data: $CB \xrightarrow{} CA$ and $CG \xrightarrow{} TG$, whereas for Nanopore it’s $TAG \xrightarrow{} TGG$ and $TAC \xrightarrow{} TGC$, and for Illumina sequencing, it is $GGG \xrightarrow{} GGT$. Fig.~(\ref{MESA_GC_graph}) and (\ref{MESA_Kmer_graph}) show the graphical representation of these points.

\subsubsection{Homopolymer and Motifs}
A homopolymer is a subsequence in which all the nucleotides are same (i.e. AAA, TTTTT). It causes hindrance in sequencing because it is difficult to read a number of repetitive nucleotides further resulting in IDS errors. Fig.~(\ref{MESA_homopolymer_graph}) further depicts this statement.

Motifs are short, repetitive patterns of nucleotide sequences (eg. TATAAA, TTGACA, etc). These types of motifs may serve several roles in chemical reactions during synthesis, and sequencing. Some of which may be beneficial while others may not. There are 58 undesired motifs in Mesa.

\subsubsection{Algorithms}
\begin{figure*}
	\centering
	\includegraphics[scale=0.70]{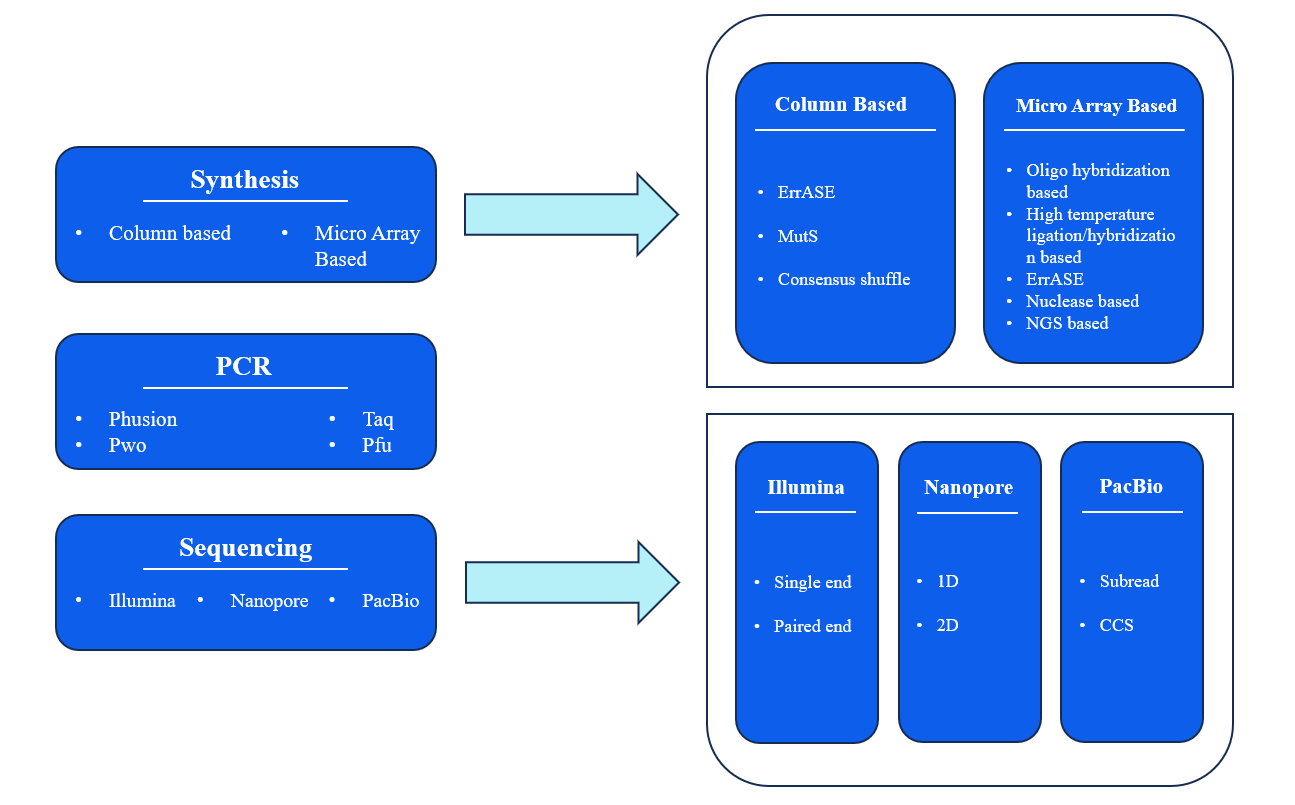}
\caption{Available algorithms in MESA software. MESA has included 2 types of synthesis, 4 types of PCR, and 3 types of sequencing algorithms.}
\label{MESAalgorithms}
\end{figure*}

Fig.~({\ref{MESAalgorithms}}) represents all the available algorithms in MESA software.

\subsubsection{MESA Customization}
\begin{figure*}
	\centering
	\includegraphics[width=\linewidth]{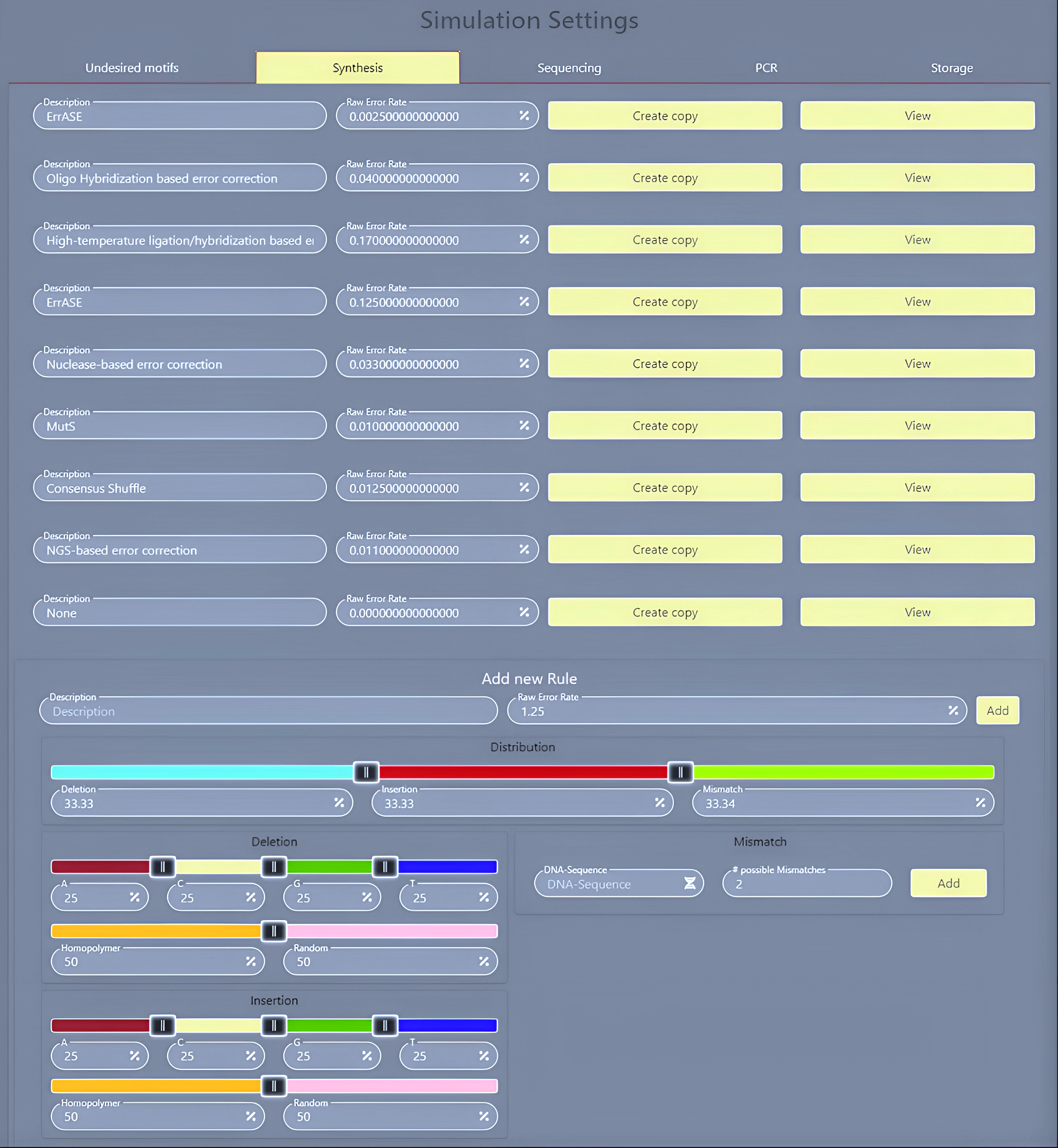}
\caption{Customization window of MESA, screenshot taken from MESA software developed by Schwarz et al. (image further enhanced for better visibility) \cite{mesamainsoftware}. © Authors, Reprinted with permission.}
\label{MESA_customization}
\end{figure*}

Here, changes can be made in the error probability graphs for the GC window, Kmer window, and Homopolymer. In the undesired motifs section, new motifs can be added and can also set their error probabilities. Changes can be made in error probabilities of various existing synthesis methods and can also introduce a new synthesis algorithm with different error probabilities. New distribution standards for the percentage of insertion, deletion, and substitution, and the percentage of ACGT in case of deletion and insertion can be made. Explicit mentioning can be done, if IDS errors occurred because of Homopolymers, or if they were random. Now the same process follows for Sequencing, PCR, and storage. Fig.~(\ref{MESA_customization}) shows the customization interface.

\subsubsection{MESA Features}
\begin{figure*}
	\centering
	\includegraphics[width=\linewidth]{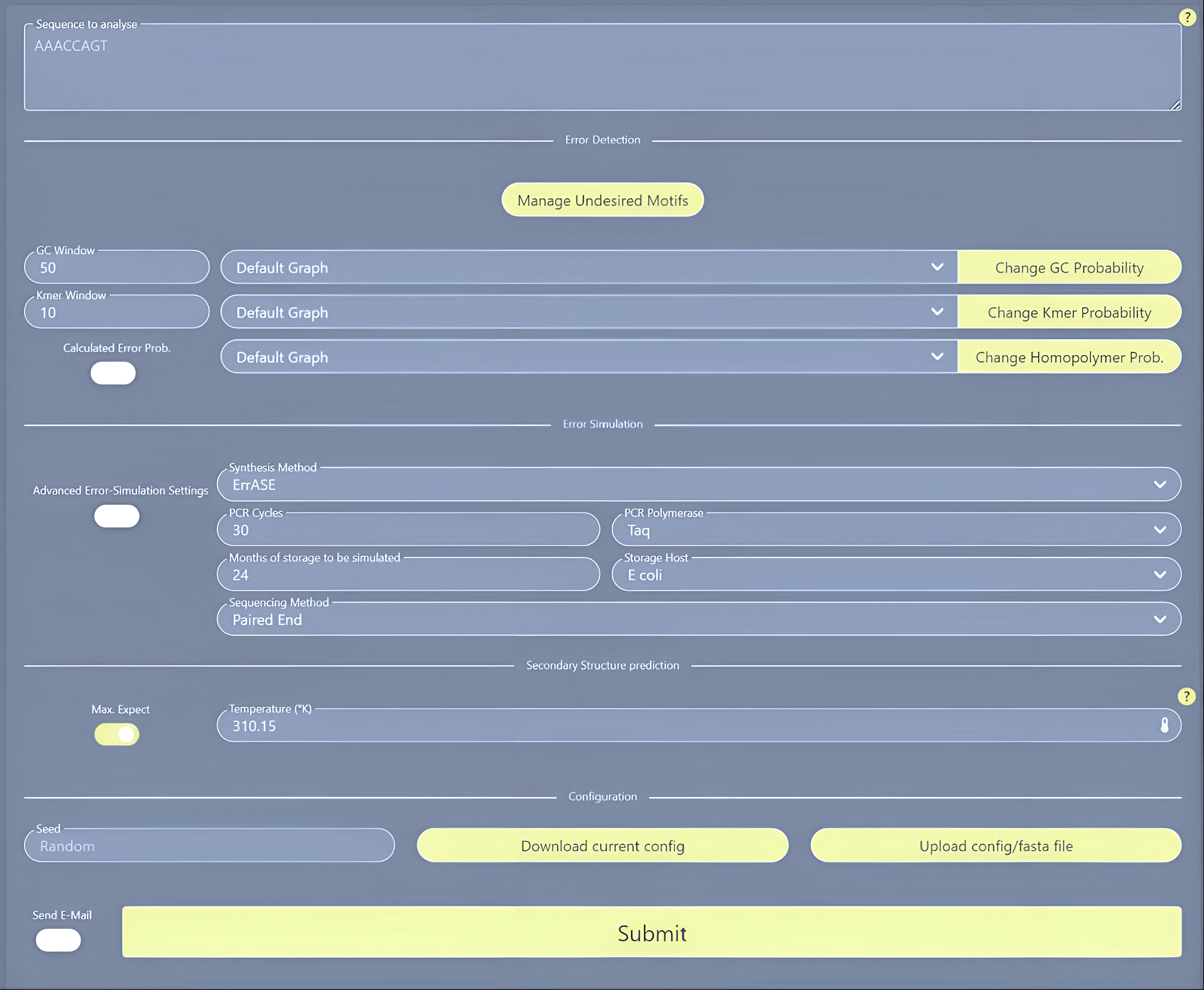}
\caption{MESA default window, screenshot taken from MESA software developed by Schwarz et al. (image further enhanced for better visibility) \cite{mesamainsoftware}. © Authors, Reprinted with permission.}
\label{MESA_default_window}
\end{figure*}

\begin{figure*}
	\centering
	\includegraphics[width=\linewidth]{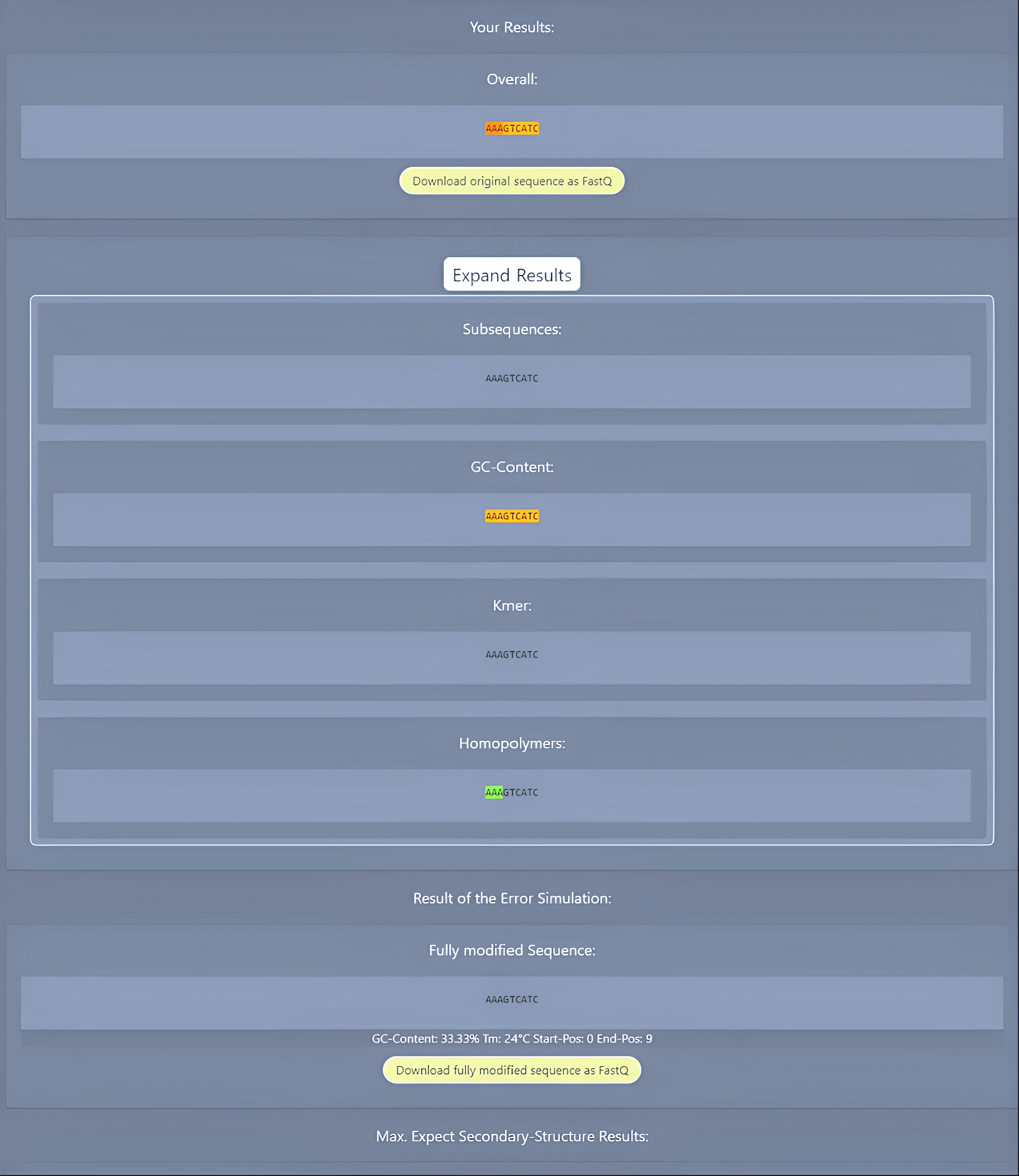}
\caption{MESA output window, screenshot taken from MESA software developed by Schwarz et al. (image further enhanced for better visibility) \cite{mesamainsoftware}. © Authors, Reprinted with permission.}
\label{MESA_output_window}
\end{figure*}

It takes .txt, and .fasta as input, and can be typed manually. On getting output, the generated sequence can be downloaded in a .fasta file. Can also give the configuration file as .txt or .fasta. In the output strand, software will provide GC content, temperature, Start pos, and End pos for any selected subsequence. Color code in output :
\begin{enumerate}
    \item After generation of output there’ll be some highlighted text in the Input strand, Subsequences, GC-content, Kmer, and Homopolymer. The color of this highlighted text will vary from light green to red. Light green represents the lowest error probability (of IDS) while red represents the highest error probability (i.e. 100\%) and gradually increases from light green to red.
    \item For modified sequence, it shows different color codes based on synthesis, sequencing, PCR, and storage IDS errors.
\end{enumerate}
Fig.~({\ref{MESA_default_window}}) and ({\ref{MESA_output_window}}) show the default window/the input window, and the output window respectively.

\subsubsection{Improvements}
This software does not include clustering, reconstruction, encoding and decoding. So, their error probabilities are not included. One can add those in order to make the simulator more realistic. Whenever larger output files are given, the website lags and even crashes, which makes it difficult to analyze the output.

\subsection{DeepSimulator}
A total of 20 sequencing simulation software exist in the market. Where Illumina, PacBio, and Nanopore are the major techniques used in DNA sequencing. There are only 4 Nanopore sequencing simulation software in the market, namely ReadSim, SiLiCo, NanoSim, and DeepSim \cite{li2018deepsimulator}. DeepSim is the only context-dependent model whereas others are all context-independent \footnote{Context-independent means in signal generation module 5-mer length is taken as default, whereas it is not the case with context-dependent}. Nanopore sequencing technology is divided into 3 parts: Sample preparation, Signal collection, and Basecalling. Their explanations are as follows:

\subsubsection{Sample Preaparation}
First of all data is converted from Binary to ACGT, and from there it will be synthesized into test tubes and stored in different bacteria. This test-tube sample will be used for our Sample preparation stage.

\subsubsection{Signal Collection}
The sampled strands will be passed onto a chosen membrane. The membrane is made up of protein or solid-state material and have nanopores embedded into it. Maximum 1 DNA strand can pass through the pore at a time. High voltage will be provided to the membrane so that the ionic current will be generated across it. So, whenever a DNA strand passes through it, it disrupts the ionic flow and a graph of fluctuating current will be generated. Now this fluctuating current will be stored in a .fasta file and it’ll be forwarded to a basecaller. MinION is the device that does this process.

\subsubsection{Basecalling}
Albacore is used for base-calling. It maps the generated fluctuating current to a reference genome (i.e. ACGT) with the Minimap method. There are other methods for mapping too, which are Graphmap, and MashMap2. Guppy is also used for base-calling. The duo goes as, MinKNOWN-Guppy, and MinION-Albacore. By doing this one can complete the sequencing process. Also, it's worth noting that it is a PCR-free process, so there’ll be no need for clustering in the future.

\subsubsection{DeepSimulator simulation in context to real life simulation}
DeepSimulator is based on Nanopore sequencing technology (see Fig.~({\ref{Simulatorsblockdiagram}}) for its high level block diagram), which tries to mimic the exact analogy mentioned above. It accepts .fasta file as an input and produces .fasta file as an output. DeepSimulator has 3 modules: Sequence generator, Signal generation module (has two parts: Pore-model component and Signal-simulation module), and Basecallers.

\subsubsection{Sequence generator}
It takes entire genome or assembled contigs (continuous DNA sequences) and coverage requirement as input. From that, it’ll generate short sequences that fulfill the coverage criterion (generally 5-mer). This process occurs in the preprocessing stage of the DeepSimulator.

\subsubsection{Pore model component}
The pore model maps the given k-mer (5-6mer) strands to a current signal. A fluctuating current graph of the corresponding k-mer strands is generated. The generated current signal is passed to the Signal simulation module. This is the module that makes DeepSim different from other software because it’s context-dependent and doesn’t always take 5-mer as a generality. It uses a deep neural network (DCTW with Bi-LSTm) to solve this problem and uses variable length k-mer depending on the situation. The main formulation of this module: given T nucleotides, a current signal will be generated of length T-4, given a 5-mer nucleotide sequence. Eg. Given a sequence ACCGTGGACT, the current signal will be f(ACCGT), f(CCGTG), f(CGTGG), f(GTGGA), f(TGGAC), f(GGACT) (i.e, the window is shifted by one nucleotide). This f() will be a deep learning model, which has 3 parts: feature representation, neural network architecture, and a generic framework for deep canonical time warping. This pore model is trained on the Pandoraea pnomenusa dataset.

\subsubsection{Problems of Pore model component}
\begin{enumerate}
    \item \textbf{Scale difference:}
Considering the frequency of the current signal - 4000 Hz, and that of the DNA sequence - 450 base/s. The scale of the input DNA strand sequence and the scale of the output current will differ. To solve this problem a Signal simulation module was introduced. 
    \item \textbf{Dimensionality difference:}
The current signal generated will be 1D but the input DNA sequence will be 4D or more, as ACGT will be “one-hot” encoded (see Table~({\ref{mappingTable}}), \textbf{“one-hot”} encoding is done in order to preserve original information).
    \item \textbf{Complex non-linear correlation:}
The conversion from input DNA strands to output current graph occurs in a noisy environment. This makes the relation between input and output rather non-linear.
\end{enumerate}
Scale difference problem can be solved by the process that will be discussed in the Signal simulation module.

\begin{table*}[]
\centering
\caption{Mapping of DNA sequence with current signal.}
\label{mappingTable}
\begin{tabular}{|c|c|}
\hline
Quaternary symbols    & One-hot encoding \\ \hline \hline
A          &      (1,0,0,0)        \\  \hline
C          &      (0,1,0,0)        \\  \hline
G          &      (0,0,1,0)        \\  \hline
T          &      (0,0,0,1)        \\  \hline
\end{tabular}
\end{table*}

\subsubsection{Signal simulation module}
The signal simulation module repeats any particular instance of the current signal at various times, corresponding to a particular base (i.e. ACGT). It also adds IDS errors at this stage.

\subsubsection{Reason Behind Signal simulation module}
Current flows at 4000 Hz through the membrane, and the nucleotide sequence bases pass through the membrane at 450 bases/s. So capturing speed of the current is 8-10 times faster than that of DNA. As a result of this, any single DNA’s current graph (i.e. amplitude) will be captured multiple times leading to repetitions in the current graph. Also, note that all the repetitions may not have the same amplitude, therefore they manually introduce IDS errors in the simulation (with the help of signal simulation module) to mimic this.

\subsubsection{Basecalling}
Albacore/Guppy will be used to map the generated current graph (by the Signal simulation module) with the ACGT genome sequence. It will use MiniMap to do this process.

\subsubsection{DeepSimulator vs Real-life simulation}
Interested readers refer to \cite{li2018deepsimulator}, Sec. 1, for comparing the real life simulation with workflow of the Deepsimulator.

\subsubsection{DeepSimulator simulations}
There are three simulation commands, which should be executed properly in order to obtain output from Deepsimulator software \cite{deepsimulatormainsoftware}. Those commands and their execution time in our VM are provided below:

\begin{verbatim}
1) "./case\_study.sh -f example/artificial
\_human\_chr22.fasta" - 1:40 min
2) "./deep\_simulator.sh -i example
/001c577a-a502-43ef-926a-b883f94d157b.true
\_fasta -n -1" - 5s
3) "./deep\_simulator.sh -i example
/artificial\_human\_chr22.fasta" - 22s
\end{verbatim}





\subsubsection{Simulation with Guppy CPU}


While running the deep simulator with Guppy CPU it is running smoothly which can be verified with task manager. The command for Guppy CPU is ./deep\_simulator.sh -i example/artificial\_human\_chr22.fasta -B 2 \cite{deepsimulatormainsoftware}.

\subsubsection{Case study simulation errors}

Running the case study file would many times end with memory error which would not be shown by the Operating system. This would be very significant when we are doing it on a Virtual machine. It would be more convenient on WSL because memory dynamic allocation would become much easier. In the script, the environment is not properly managed. There are some errors regarding deactivating.

\subsubsection{Problems in DeepSimulator}

The Guppy-GPU and Guppy-CPU (in VM) are not working because of the absence of a shared library file. Also, the code is running in older Python versions. Some .sh files are also not working because of old dependencies. Executions of the sample codes resulted in the output which got stacked on top of each other in the same file and folder as before. For some sequences the divergence between the raw simulated signal and the real signal is large. There are many computational challenges. The installation process is very complex and interpretation is also hard because it’s not GUI-based.


\section{Comparison}\label{sec:4}

Storalator is the easiest to use in comparison to Mesa and DeepSimulator. Only takes a .txt file containing ACGT strands and not a fasta file. Latest processes such as synthesis, PCR, sequencing, clustering, and reconstruction algorithms are included. We can do changes in error simulation as per our requirement. It's a windows software and we can directly run it with an application file. Results are not highly accurate, because it cover so many domains. Also lastly it doesn't cover encoding, decoding, and storage processes.

Mesa is easy to use in comparison to DeepSimulator, but a bit difficult than Storalator. Takes fasta file containing ACGT strands as input. Processes such as synthesis, PCR, storage, and sequencing are included. There are additional features as compared to storalator, such as PCR adjustment, storage host, time for storage, and temperature. We can do changes in error simulation as per our requirement, along with more flexibility. It's a website where we can directly give our input and it'll do the simulation. Results are more accurate than Storalator, but less than DeepSimulator. Also lastly it doesn't cover encoding, decoding, clustering, and reconstruction.

DeepSimulator is very complicated to use. Implements a deep learning model on fasta files which takes ACGT strands as input. Can do changes in error simulation, but has got less flexibility. Only includes sequencing and coverage (PCR) that to Nanopore in particular. It's a software that we can run on a Linux machine and we can give our input as a fasta file into the terminal. Results are more accurate than both Storalator and Mesa, considering it's highly domain specific. Also lastly it doesn't cover encoding, decoding, synthesis, storage, clustering, and reconstruction.

Crisp comparison between all three DNA storage simulators are given in Table~({\ref{comparisonTable}}). It includes comparison between all the basic DNA storage processes, along with easy to use from user perspective, and finally the accuracy.

\begin{table*}[]
\centering
\caption{Comparison between all three Simulators. It includes all the basic DNA storage processes, along with easy to use from user perspective, and finally the accuracy.}
\label{comparisonTable}
\begin{tabular}{|c|c|c|c|c|c|c|c|c|c|c|c|c|}
\hline
Simulators    & Easy to use & Input  & Encoding & Synthesis & Storage & PCR & Sequencing & Clustering & Reconstruction & Decoding & Output & Accuracy \\ \hline \hline
Storalator    & High        & .txt   & No       & Yes       & No      & Yes & Yes        & Yes        & Yes            & No       & .txt   & Low      \\ \hline
Mesa          & Moderate    & .fasta & No       & Yes       & Yes     & Yes & Yes        & No         & No             & No       & .fasta & Moderate \\ \hline
Deepsimulator & Low         & .fasta & No       & No       & No      & Yes  & Yes         & No         & No             & No       & .fasta & High  \\  \hline
\end{tabular}
\end{table*}


\section{Technology Limitations}\label{sec:5}
Seeing real-life's point of view, a lot more research can be done for storing data in DNA at a lower cost, as well as increasing the read and write speed of DNA. From simulators point of view new algorithms can be made, and more precise solution can be used in order to make these simulators more and more realistic.


\section{Future Scope: Exploration on JPEG DNA}\label{sec:6}
In digital images archiving and storage, JPEG standards are used widely now-days. Therefore, with an effective image coding format to generate synthetic DNA molecules, the JPEG Committee is expected to be able to address the issues of DNA-based storage. In order to initiate a standardization effort, JPEG DNA was formed as an exploratory activity inside the JPEG Committee to examine use cases, identify needs, and evaluate the state of technology in DNA storage for the goal of picture archival using DNA \cite{jpegdna}. Multiple workshops were organized for this purpose.


\section{Conclusion}\label{sec:7}
To conclude this paper, we are providing some information regarding appropriate usages of these softwares according to user requirements. Storalator is very user friendly and has good GUI along with good combination of different algorithms. MESA is also user friendly and has more flexibility in terms of features. DeepSimulator is not so user friendly, but is highly domain specific, and has less flexibility amongst its features, but at the end it will provide very accurate synthesis simulation.

But always there is more scope for improvements in all the existing software. Binary to ACGT encoding and decoding can be incorporated. Softwares can be made more user-friendly so that everyone can use and understand it. More and better algorithms/solutions can be incorporated. Development of DNA simulators can be said to be in a very early stage of development and there is a lot of room for improvement by incorporating more domains and algorithms.

 \section{Acknowledgment}
The authors would like to thank  Gadi Chaykin, Omer Sabary, Eitan Yaakobi and Dominik Heider for reading the first draft of the paper and providing useful feedback and giving permission to use the figures. The authors also thank Hem Bhalodiya and Dev Patel for useful discussions. 

\bibliographystyle{IEEEtran} 
\bibliography{dnastorageeref}

\end{document}